# A Brief Review of Single Event Burnout Failure Mechanisms and Design Tolerances of Silicon Carbide MOSFETs

Christopher A. Grome[1,a], Wei Ji[1,b]

[1]Department of Mechanical, Aerospace and Nuclear Engineering, Rensselaer Polytechnic Institute, Troy, New York, 12180, U.S.A.

[a]grome1@llnl.gov, [b]jiw2@rpi.edu

**Abstract:** Radiation hardening of the MOSFET is of the highest priority for sustaining high-power systems in the space radiation environment. SiC-based power electronics are being looked at as a strong alternative for high power spaceborne power electronic systems. The SiC MOSFET has been shown to be most prone to SEB of the radiation effects. The knowledge of SiC MOSFET device degradation and failure mechanisms are reviewed. Additionally, the viability of rad-tolerant SiC MOSFET designs and the methods of SEB simulation are evaluated. A merit system is proposed to consider the performance of radiation tolerance and nominal electrical performance. Criteria needed for high-fidelity SEB simulation are also reviewed. This paper stands as a necessary analytical review to intercede the development of radiation-hardened power devices for space and extreme environment applications.

## I. Introduction

Due to the aggressive push in the electronic systems market for a smaller, better, cheaper, faster philosophy, microelectronics manufactured by commercial processes are increasingly being used. Commercial advancements in developing smaller feature size has led to increased component density in power systems, drastically increasing these advanced systems' radiation sensitivity. Utilizing these dense power systems that are commercially developed adds noticeable risk if the electronics on board are not "hardened" to the radiation environment. A strong interest in the power electronics market has been shifted towards wide-bandgap (WBG) semiconductors. The larger energy gap intrinsic to WBG materials permits power devices to operate at much higher voltages, frequencies and temperatures compared to the traditional semiconductor materials like Silicon (Si) and Gallium Arsenide (GaAs). The resilient characteristics of WBG materials have made them strong candidates for developing a radiation hardened power device solution. Although it has been shown that WBG power devices have better radiation tolerance than conventional semiconductor counterparts, radiation-induced reliability issues are still observed at high operating voltages that leave this technology incomplete [1].

With the current ambitions and innovations by industry, the demand for efficient, radiation-hardened and thermally resistant power devices are crucial for sustaining high-power systems in high-radiation environments. The design of advanced radiation tolerant electronics has boundless use in the extreme environments that exist both terrestrially and in outer space. Due to the rapid growth in the commercial high-voltage WBG-based power electronics, rad-hard WBG technology has the potential to transition into the mainstream radiation-hardened power market in space.

Advances in the application of a variety of semiconductor materials, processing technologies, and design techniques have opened unprecedented opportunities for enhancement of space electronics in the areas of power management, surveillance, on-board computation, and communication. A development such like this comes at an imperative time as NASA plans to start a sustained human presence on the Moon. During demanding missions, this rad-hard technology can deliver infrastructure power to environments this previously was not available, vastly benefitting lunar exploration capabilities. Compared to the current Si-based power management and distribution systems (PMAD), the implementation of WBG power components have been shown to add triple the amount of voltage and reduce the system power losses by more than 50%. Due to WBG power components high size, weight and power (SWaP) potential, there is less need for more Si-based devices, shown to save an estimated 20% in mass and volume [2]. By satisfying multiple requirements with an optimized PMAD system, there is now more space and mass available for additional functionality, better efficiency and greater energy production. Additionally, as electrically powered propulsion systems are becoming more popular for deep space propellant systems and new desire for airplanes, an advancement in rad-hard electronics would greatly benefit the power needs for these concepts. By providing high power conversion for all parts needed in the power distribution system, an electrically propelled vehicle can perform at a much more desired capability [3]. This opens the gateway to a completely new method of propulsion that is imperative for a new age of travel and deep space exploration.

Terrestrially, the development of radiation tolerant electronics would provide instrumentation for monitoring advanced nuclear reactor systems. Providing novel sensors to measure the structural health of reactors, maintenance and system reliability can be considerably optimized. Moreover, military and air force systems can be bolstered from radiation exposure. With the emergence of neuromorphic computational paradigms of new device and circuit architectures and of specific device structures and coupled oscillators as computational elements, there is a new frontier to be explored systematically. These architectures and algorithms are highly dependent on radiation sensitivity performance and power requirements. The security and reliability of all these emerging technologies would drastically benefit from advancements in radiation-hardened power systems.

The use of silicon carbide (SiC), a WBG material with higher critical electric field and thermal conductivity characteristics than traditional silicon (Si), has been thoroughly studied to improve radiation susceptibility in power devices. A comparison of the two materials summarizing their electrical and thermal properties can be seen in Table 1 [4]. SiC device fabrication and techniques have been thoroughly studied over the past few decades, bolstering the technology readiness and maturity of this material compared to more novel researched WBG semiconductors. Although SiC may not be the superior semiconductor in the future, advancing the understanding of the material and SiC-based devices will provide techniques that will benefit the development of advanced WBG-based devices for the future. The Junction Barrier Schottky (JBS) diode and Metal Oxide Semiconductor Field Effect Transistor (MOSFET) have been identified as the two most critical power devices for highpower applications [3]. Experimental data has shown that SiC power devices are most susceptible to the single event effects (SEEs) caused by heavy ions [5], [6], [7] and neutrons [8], [9], primarily failing due to single event burnout (SEB). Although SiC devices perform more optimally at higher voltages than Si-based devices under irradiation, the threshold voltage the devices experience SEE is still undesirably around 40% of the device's rated operating voltage [10]. Many experimental and simulation studies have been done in attempt to improve this percentage, however there is yet to be a device that fulfills the high voltage rad-hard need of industry commercially available. In order to advance the limit of this current technology a fundamental understanding of the device's failure mechanisms need to be reviewed.

| Property | Si | 4H-SiC |
|---|---|---|
| Bandgap, $E_g$ (eV) | 1.12 | 3.26 |
| Dielectric constant, $\varepsilon_r$ | 11.9 | 10.1 |
| Electric breakdown field, $E_c$ (kV/cm) | 300 | 2,200 |
| Electron mobility, $\mu_n$ ($cm^2$/V-s) | 1,500 | 1,000 |
| Hole mobility, $\mu_p$ ($cm^2$/V-s) | 600 | 115 |
| Thermal conductivity, $\lambda$ (W/cm-K) | 1.5 | 4.9 |

Table 1 Comparison of Si and SiC

**Motivation and Objective**

Studies have proposed numerous theories for the MOSFET failure mechanisms each with their own respective reasons. Ion bombardment experiments on Si Bipolar Junction Transistors (BJTs) shown in [11] demonstrated that the generated charge carriers flowing through the P-base region create a voltage drop that can turn on the transistor. This phenomenon is known as a BJT turnon and is a mechanism that causes a localized positive feedback current inducing the generation of carriers eventually leading to thermal runaway and destruction of the device. Experimental and simulated studies done in the 1990's revealed that the Si MOSFET burnouts were induced by the turn on of the parasitic BJT inherent in the MOSFET design [12], [13], [14], [15], [16] and this claim has since been supported. The Si diode, where a BJT structure is not present in the design, has been studied under ion bombardment to evaluate its SEB threshold voltage. It's been shown that Si Schottky can survive heavy ion strikes up to 100% of its rated voltage and can operate reliably at 75% of the rated breakdown [17]. However, the Si MOSFET at LETs greater than 15 MeV-cm2 /mg experiences failure at or below 40% of its rated voltage [1]. This contrast in SEB threshold adds complimentary support to the devices differing failure mechanisms: turn-on of the BJT in the Si MOSFET and a plasma induced joule heating thermal breakdown in the Si diode.

It was commonly assumed that the BJT turn on mechanism would be responsible for SEB in SiC-based MOSFETs. Differing from experimental studies on Si JBS and MOSFETs where the SEB failure and degradation thresholds are starkly different, experimental studies on SiC JBS and SiC MOSFETs are almost exactly the same [1], [18], [19]. The similarity in burnout characteristics suggests a common SEB failure mechanism between the devices. Shoji et al. [20] was one of the first to propose a failure mechanism opposing the conventional ideas previously documented for the SiC MOSFET. The similar failure characteristics between the SiC JBS (does not have parasitic BJT) and SiC MOSFET (does have parasitic BJT) support the claims made and since then many works have come forward to support [6], [21], [22], [23], [24] and refute the parasitic BJT [4], [18], [20], [25], [26], [27], [28] as the regenerative failure mechanism responsible for SEB in SiC MOSFET.

A comprehensive review of these theories is necessary to elucidate discrepancies and further the understanding of WBG power devices to move this technology further. While SEB failure mechanisms in diodes are relatively agreed upon, the mechanisms that corrupt SiC MOSFETs are still debated and will be the primary scope of this paper. Similarly, an analytical review of proposed simulated rad-hard MOSFET designs is necessary to intercede the practicality of development. The desire by industry requiring the most research and future work are fully efficient MOSFETs that can survive catastrophic SEEs up to 1200V [3]; this 1200V burnout mark will be a standard when analyzing and attributing merit to rad-hard designs. This paper presents a review of these design models as proposed throughout history, explaining their merits, limitations, similarities and differences. The objective of this paper is to give current developers a transparent idea of the SEE failure mechanisms effecting SiC power devices and the suitability of the proposed rad-hard designs to help lead to the commercialization of devices that meet the performance standards of industry.

This paper is organized, first, discussing the evidence provided for the SEB failure mechanisms chronologically proposed for Si and SiC power diodes and MOSFETs. Section III compares the radiation tolerances of different SiC MOSFET structures and assesses the merit of the rad-hardened counterparts. In Section IV, the conclusion discusses the current state of the technology readiness and work needed in the future to bring radiation tolerant devices to market. Lastly, Section IV reviews the background for high-fidelity simulations, failure criteria and discusses future simulation developments to accurately replicate heavy ion strikes on rad and non-rad hardened devices.

## II. Review of Proposed SEB Failure Mechanisms

Unexpected failures and reliability issues were observed in high power silicon devices that had expected lifetimes of over 25 years early after their development [29]. The effect of the SEB was fist observed in 1986 by Aerospace Corp who reported a destructive “latched current” effect in several n-type MOSFETs [14]. Experimental studies determined a correlation between failure rate and high DC voltage. It was not long after until a group discovered these failures stopped happening when tested 140 meters below ground. Kabza *et al.* [29] attributed these failures due to the terrestrial cosmic radiation that is experienced on Earth. Throughout the late 1980s and 1990s, many groups employed experimental and simulation radiation studies on Si power devices to determine the failure mechanisms responsible for SEB.

### Evidence of the BJT Turn on in Si MOSFETs During SEB

Experimental studies on the Si Bipolar Junction Transistor (BJT) revealed that heavy ion radiation could parasitically turn on the transistor while it is in its blocking off state. Johnson [12] was of the first to describe the physical mechanisms behind this phenomena. Upon ion strike, electron-hole pairs are generated along its track length. This ionization creates a plasma filament able to support a short lived current source where holes flow towards the lateral base region and the electrons flow toward the collector. This hole current forward biases the base-emitter junction surpassing the threshold voltage required to turn on the BJT. Once the BJT is turned on, the currents within the device will regeneratively increase until the simultaneous high current and high voltage trigger second breakdown leading to thermal destruction of the device. This mechanism was experimentally validated by implementing an inline resistor between the collector lead. The addition of a resistor here would suppress the regenerative feeding of current in the turned on BJT and would thus prevent the reaching of second breakdown. Johnson showed this addition prevented burnout in all cases.

Due to the parasitic BJT structure that is inherent to the MOSFET design, experimental and simulation studies determined that the BJT failure mechanism was also responsible for the Si MOSFET SEB [12], [13], [14], [15], [16], [30], [31], [32], [33], [34], [35], [36], [37], [38]. Kuboyama *et al.* [15], showed enhanced charge collection due to the turn on of the BJT flooding electrons from the n+ source in the Si MOSFET. This result was compared to a Si MOSFET that was fabricated without this n+ source, resembling the structure of a diode. The enhanced charge collection was not seen in this device and as a result had better SEB tolerance. The result indicated the need for the parasitic BJT to trigger SEB and verified its enhanced current effect. It was shown in [39], [40] that limiting current with a series resistor also prevented burnout in the Si MOSFET. When the parasitic BJT turns on it draws current from the source to regeneratively feed into the device. The addition of a resistor at the source has shown to improve SEB susceptibility in the Si MOSFET and the Si BJT, experimentally validating support of this failure mechanism. Experimental testing on the Si MOSFET revealed that the SEB threshold for this device occurs at less than 40 percent of its rated voltage (ratio is normalized voltage) [1]. This relationship is illustrated in Fig. 1 in a Si-SiC experimental comparison.

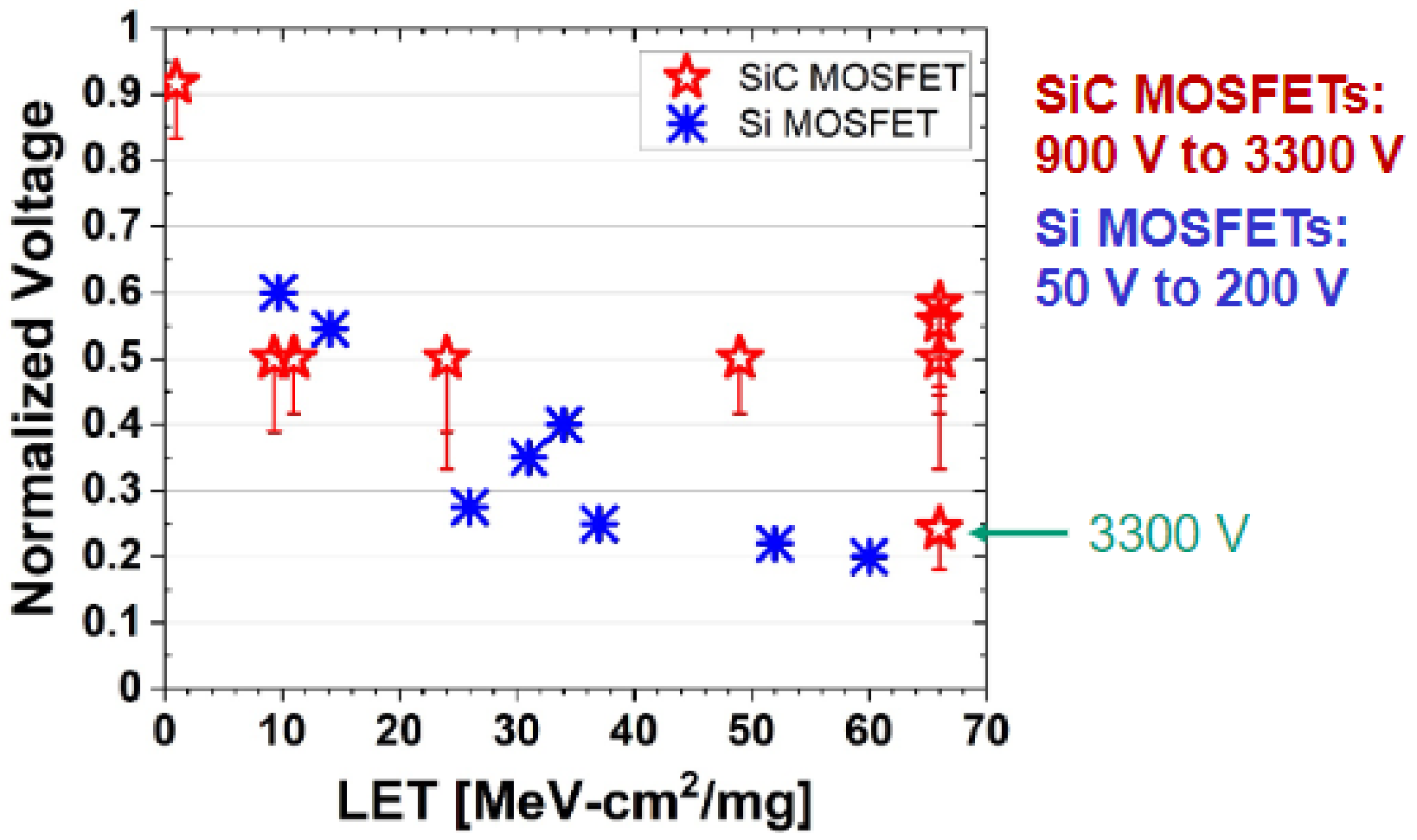

Fig. 1 SEB threshold comparison Si vs SiC MOSFET. After [1].

Since Si diodes lacked the parasitic BJT structure in its design, these devices (other than high voltage designs) were thought to be immune to destructive SEEs [36]. Charge collection studies conducted on the high voltage PiN Si diodes observed charge amplification when voltage at the cathode is increased due to an increase in internal electric field. This effect was reported as the charge multiplication phenomenon and was used to explain the failure in high voltage diodes [17], [41], [42]. Since this failure mechanism was dependent upon the application of a high voltage, the low voltage diodes remained to be considered immune to destructive SEEs. In 2012, an experimental study testing Schottky diodes DC-DC conversion failed due to SEB under heavy ion bombardment. These diodes failed at 50 percent of their rated voltage including a diode rated at 45 V [43]. It was revealed the Schottky diode was more susceptible to SEEs because of their lack of protection at the Si/metal interface, causing melting at this area from a shorting between the anode and cathode. This failure phenomena was not observed in diodes that protected the metal/Si interface [17]. The generated current from a heavy ion caused melting at this interface shorting the anode and the cathode leading to premature failure. More recent experimental and simulation studies have described the failure mechanisms for the Si PiN and Si JBS diodes [1], [8]. Lauenstein *et al.* deemed impact ionization insufficient for Si PiN diode SEB. During ion strike, impact ionization leads to localized heating at the epitaxial (epi) /substrate interface that causes a thermal generation of intrinsic carriers. The increase in intrinsic carrier concentration overwhelms the background doping in the device, leading to more local heating creating a thermal runaway feedback loop until the device is thermally destructed. This is described in Eq. 1 illustrating the intrinsic carrier concentration as a function of temperature.

$$n_i(T) = \sqrt{N_c(T)N_V(T)}\exp\left(-\frac{E_g(T)}{2kT}\right) \quad (1)$$

As temperature increases, the term $E_g(T)$, which is the bandgap of the material, decreases, however the terms $N_C(T)$, describes the conduction density-of-state, and $N_V(T)$, describes the valence density-of-state, increases. The result causes the intrinsic density, $n_i$, to increase with increasing temperature eventually reaching the intrinsic temperature. This is the point where the intrinsic density equals the background doping concentration and marks the beginning of the thermal instability process. The increase in intrinsic carriers becomes regenerative since this increases current density leading to further heating of the lattice leading to thermal generation of even more carriers. This process continues until the device catastrophically fails from a thermal runaway and describes the second breakdown process that power devices undergo when stressed beyond their blocking capabilities.

Although the diode is susceptible to SEEs, the Si diode has shown a greater SEB normalized threshold compared to the Si MOSFET. Lauenstein *et al.* showed 83% of Si Schottky diodes biased at 75% of its rated reverse voltage irradiated by an LET of 59 MeV-cm$^2$/mg can survive without permanent damage. Additionally, it was shown all Si Schottky diodes under the same radiation conditions can survive with no damage when biased at 50 percent of the rated reverse voltage. These data show considerable more radiation tolerance compared to the Si MOSFET in comparable conditions illustrated in Fig 4. These results correlate to the differing mechanisms effecting these two types of devices. Intuitively, this further supports the BJT turn-on as a primary failure mechanism and explains the greater susceptibility to SEB seen in the Si MOSFET. A more detailed review of the Si MOSFET SEB failure mechanism is made by Titus *et al.* [44].

**Review of SiC MOSFET SEB Failure Mechanisms**

In SiC-based MOSFETs it was commonly assumed that the BJT turn-on mechanism would be responsible for SEB. This became the conventional way of thinking and may have led to confirmation bias when investigating the device. Zhang [21], was one of the first to investigate the SiC MOSFET SEB failure mechanisms in 2006. Zhang credits the extensive Si MOSFET studies done in the 1990s as the basis for the understanding of the SEB failure mechanism in these devices. The simulations conducted were in 2D and did not verify the results with an experimental study on SiC devices, but rather were verified with previous Si MOSFET studies. The SiC MOSFET under heavy ion bombardment (of an unknown LET) showed burnout at 215 V. The device also showed burnout on the order nanoseconds, which disagrees with SiC device simulations completed in recent years [18], [20], [23], [24], [25], [26]. Although the static breakdown voltage ($BV_{pp}$) of the device was not given, the drift doping profile of $3x10^{15}$ cm$^{-3}$ was given. Using ideal breakdown methods listed in Eq. 2, the $BV_{pp}$ should be on the order of thousands.

$$BV_{pp} = 4.77 * 10^{14}\ N_D^{-5/7} \quad (2)$$

A breakdown due to SEB at 215V leaves a normalized threshold voltage of .05 or smaller which is less than any recorded threshold voltage recorded experimentally [1], [17]. The fidelity of this simulation is highly questionable. Similar to the SEB mechanisms in Si, Zhang provides evidence of the SiC MOSFET exhibiting BJT turn on. Zhang showed high sensitivity when the strike location was at the n+/p/n region. When ions were struck overtop, the parasitic BJT structure of the MOSFET burnout was triggered at lower voltage biases than striking at the channel or the p-body regions. However, in more recent experimental and simulation SiC MOSFET studies, the striking angle and location have been proven to be most susceptible directly overhead and with minimal positional dependence [4], [23], [25]. Zhang provides additional evidence for a BJT turn-on by displaying the electric field peak shifts during strike. Prior to impact, the peak electric field in the device is at the n+ source/p-base junction, however during the strike the peak shifts to the n-drift/n+ substrate interface. Zhang explains this shift as "base-pushout" due to the current injection of the parasitic BJT, which shifts carrier concentration and impact ionization to epi/substrate region. However, this "base pushout" is known as the Kirk Effect and has been observed in power devices without a parasitic npn structure [45]. SiC diodes have shown peak electric field shifts from the P+/n drift region to the n drift/n substrate in simulations [20] and may not be a unique phenomenon to the MOSFET. Zhang concludes that the failure mechanism of the SiC MOSFET is no different than the Si MOSFET and characterizes the sequence of SEB exactly as stated in Si MOSFET studies [12], [13], [14], [15], [16]. There still lies many questions about the accuracy of the simulation models employed and the logic used to advocate for the failure mechanisms concluded, making it hard to believe confirmation bias did not influence Zhang's results.

The advocacy for the BJT turn-on in SiC MOSFETs continued in 2011. Griffoni *et al.* [22] documented the first experimental neutron exposure testing on SiC power MOSFETs. Testing 1200 V rated devices, no SEB events were observed when irradiated with $1.5 \cdot 10^4$ n/cm$^2$/s and $5.7 \cdot 10^4$ n/cm$^2$/s for the 50-MeV and 80-MeV neutrons, respectively. However, the Si counterparts rated at lower voltages all experienced failure. Griffoni argued the lack of SEB in the SiC MOSFET was attributed to a larger built in potential between the n+ source and the p-base region requiring higher energy particles to turn on the parasitic BJT. The basis for Griffoni's failure mechanism understanding came from Zhang in [21].

More recent simulations investigating the physical mechanisms of SEB in these devices have also agreed with Zhang. Johnson *et al.* [23] aimed to find experimental evidence of the BJT turn-on mechanism. Johnson employed a pulse-laser two-photon absorption experiment to observe charge collection in the SiC MOSFET. The collected charge results from the MOSFET were compared against a SiC JBS diode and are displayed in Fig 2. The devices were exposed to an ionizing source while reverse-biased and their collected charge measurements were recorded. The charge amplification seen in Fig. 5 in the MOSFET suggests parasitic BJT activation, where the diode's results indicate there is no injected carrier source.

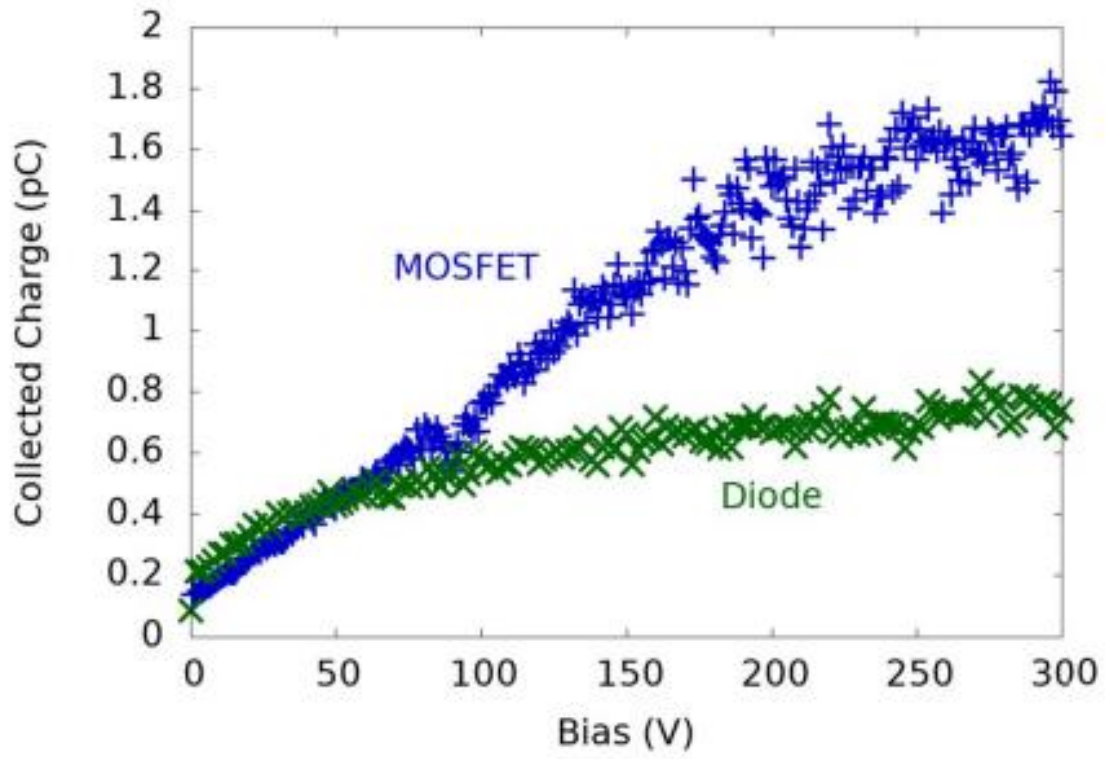

Fig. 2 Diode and MOSFET collected charge at reverse-biases of 0 to 300V. After [23].

Johnson further conducted 2D simulations of the devices under heavy ion strike to investigate the charge collection at specific regions. The results from the simulations in Fig. 3b show the diodes are independent to collected charge with respect to position, where the MOSFET in Fig. 3a exhibited significant gain in collected charge for a strike near the parasitic n+\p\n region.

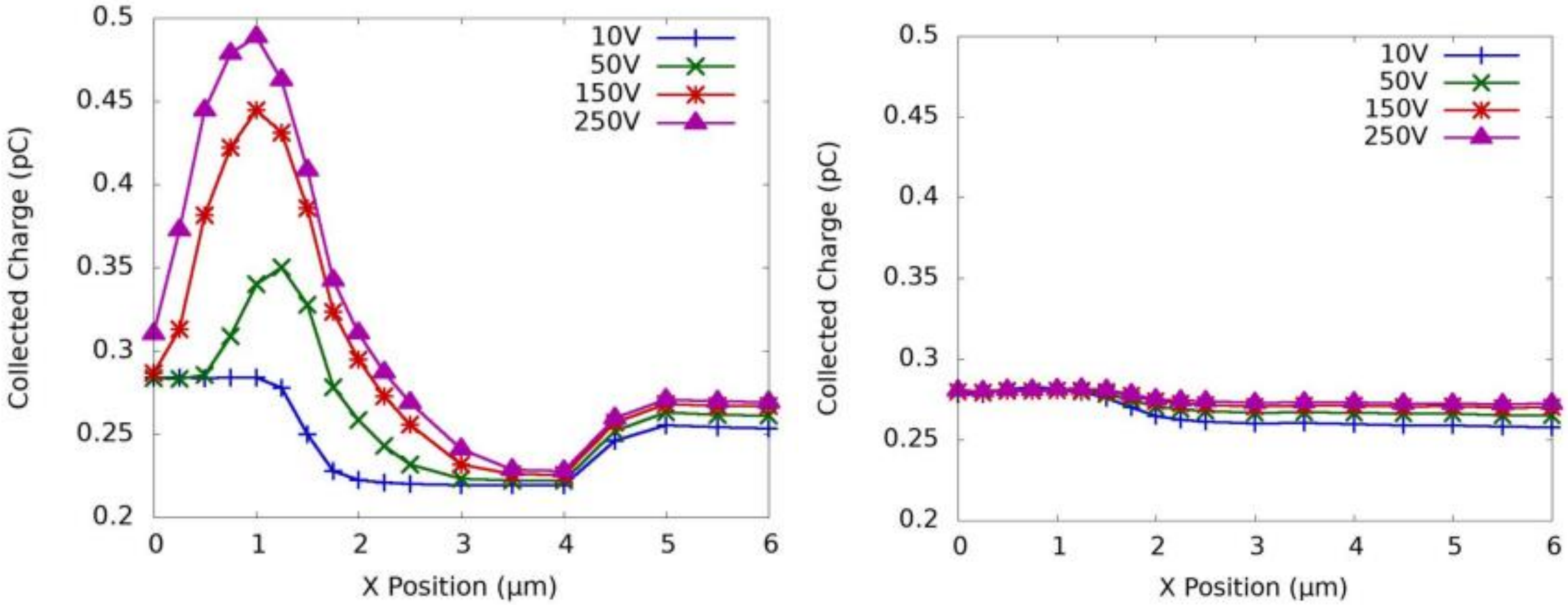

(a) (b)

Fig 3 a) MOSFET collected charge simulation results b) Diodes collected charge with respect to devices lateral position. After [23].

While the evidence of BJT activation seems overt, Johnson did not observe SEB unless impact ionization was implemented in the MOSFET simulation. The charge amplification from the n+/p/n region alone was not enough to cause thermal destruction of the device and Johnson suggests there is another complimentary mechanism responsible for SEB. Johnson concludes the parasitic bipolar amplification is an important mechanism contributing to SEB, but did not quantify the amount of contribution it provides with respect to the impact ionization. Additionally, the positional dependence observed in the simulation revealing the significant charge contribution from the n+/p/n region was not backed up by the experimental results originally examined, leaving a discrepancy between the two studies. Although charge amplification was evidently seen in the MOSFET in both simulation and experiment, the evidence it contributes toward SEB was not drawn.

Zhou *et al.* [24] compared Si and SiC MOSFETs SEB susceptibility by exploring the failure modes in each device. Drawing from previous work in [46], [47], Zhou attempted to verify the turn-on of the parasitic npn transistor in the SiC MOSFET by observing the 'base-pushout' phenomenon. Two different heavy ion strike conditions were simulated on 600V Si and SiC devices. Condition A is at $V_{ds}$ = 500 V and LET = .01 pC/μm; condition B is at $V_{ds}$ = 80 V and LET = 1 pC/μm. Electric field distributions were shown over time throughout the simulation for each condition. In condition A, the Si MOSFET experienced SEB and the SiC MOSFET does not. The Si MOSFET showed a peak electric field shift from the n+/p body interface to the n/n+ homojunction region, while the SiC MOSFET suppressed an electric field shift. Zhou postulated the reason the SiC MOSFET did not burnout in condition A, is due to the high doping profile (two orders of magnitude larger than Si) in the drift region. The low LET of the particle can be more easily suppressed because a larger number of electrons from injection are needed to trigger BJT turn-on. In condition B, the Si MOSFET did not burnout, however the SiC MOSFET did. The electric field distributions showed the same shift in peak electric field to the homojunction in the SiC MOSFET that the Si MOSFET experienced except on a picosecond timescale. Zhou claims the base-pushout phenomenon from the BJT turn-on induces a peak electric field shift known as the Kirk effect. The electric field eventually surpasses the critical electric field triggering second breakdown in the device. Since both the Si and SiC MOSFETs saw the same electric field shift, Zhou advocates the failure mechanisms are therefore the same. Since it is universally agreed upon the Si MOSFET fails via BJT turn on, Zhou explains this must be the case for SiC. As stated before, this "base pushout" shifting of the peak electric field have been observed in SiC diodes shown from P+/n drift region to the n drift/n substrate in [20] and may not be a unique phenomenon to the MOSFET. Additional evidence via carrier current density distributions need to be performed in order to show a true biasing of the parasitic BJT.

Witulski *et al.* [6] investigated the SiC MOSFET via high fidelity 3D TCAD simulation. Their approach was directed at determining the contribution of the BJT turn on mechanism during burnout. Two MOSFET models were employed in order to do so: one with nominal impact ionization and one where impact ionization was artificially turned off. The results showed that without impact ionization employed in the model the device did not experience SEB. The current densities of the two models illustrating the difference in impact are shown in Fig. 4.

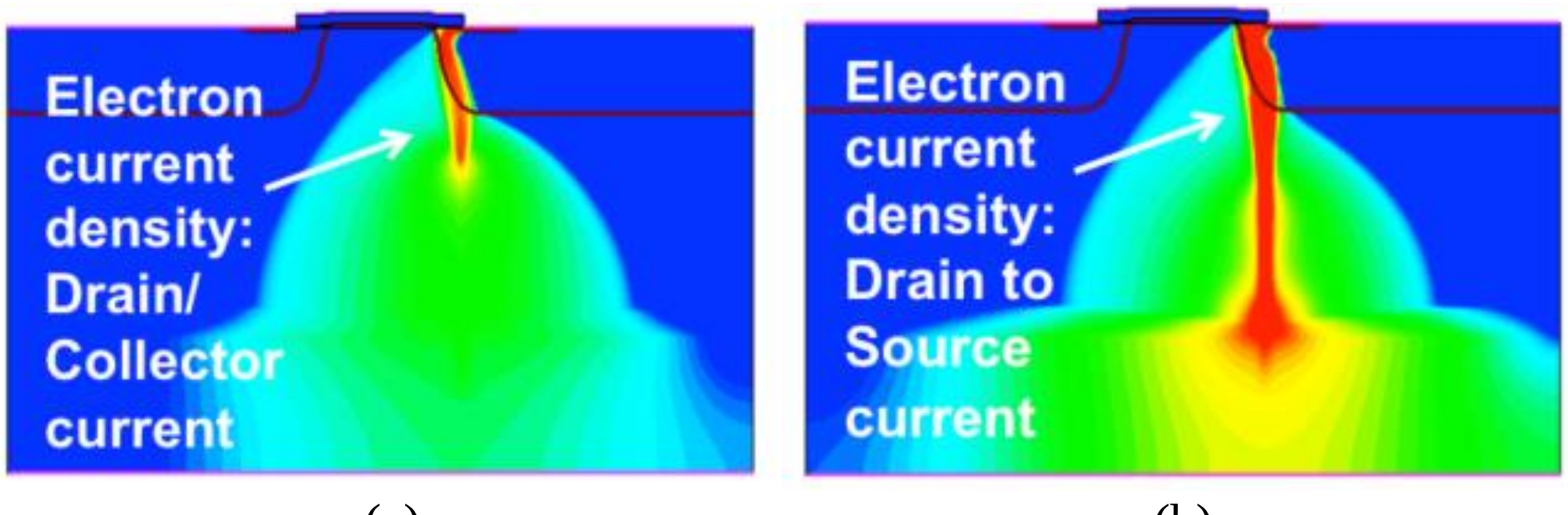

(a) (b)

Fig. 4 a) Model with no impact ionization isolating for BJT contribution b) Model with impact ionization turned on. After [6].

Peak current density was observed at the npn region in Fig 4a, however left to its independent processes could not cause second breakdown. It was recorded that SEB could only occur when impact ionization model was turned on. Witulski argues that the impact ionization at the epitaxial-drain junction work as a complimentary mechanism to the parasitic turn on of the BJT, which leads to increased drain currents that trigger SEB. What is not quantified in this study is the contribution of each mechanism. The BJT mechanism may turn on, but may not contribute an amount comparable to the impact ionization at play. It was shown when isolated, the BJT turn-on could not cause SEB, but the isolation of the impact ionization mechanism was not explored in this study.

Shoji *et al.* [20] was also interested in the contribution of the mechanisms that cause power device SEB. Similar to Kuboyama's Si study in [15], where the parasitic BJT was removed from the MOSFET to see if SEB susceptibility was effected, Shoji removed the n+ source in the SiC MOSFET. If the BJT mechanism was a considerable contributor to SEB, the removal of this region would show increased SEB tolerance. However, the results showed the device without the n+ region burned out at the exact same voltage bias as the traditional SiC MOSFET suggesting alternative mechanisms are causing SEB.

Differing from experimental studies on Si JBS and MOSFETs where the SEB failure and degradation thresholds are starkly different, experimental studies on SiC JBS and SiC MOSFETs are almost exactly the same [1], [17], [18]. In Fig 5 from [18], the alignment of the degradation in SEB thresholds of the two SiC-based devices are compared.

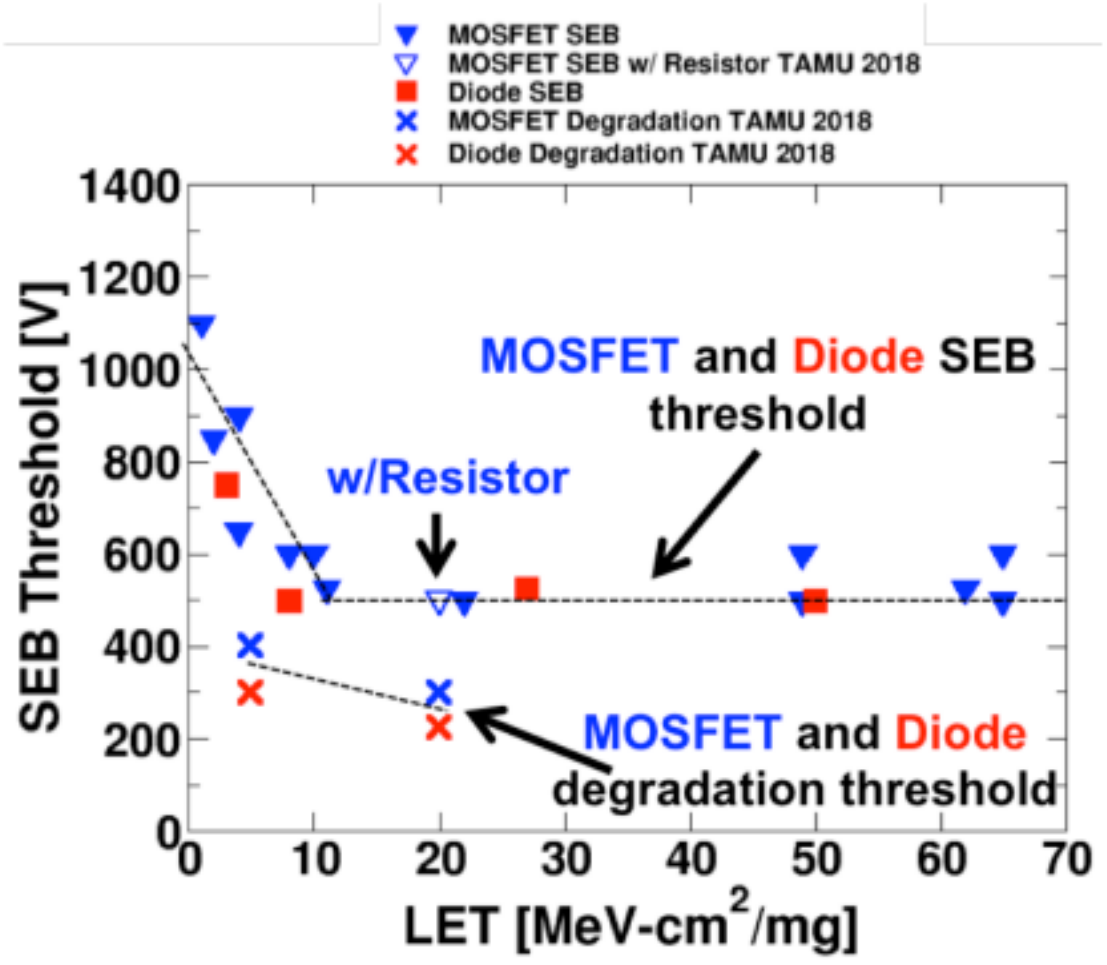

Fig. 5 Comparison of SiC MOSFET and Diode SEB threshold. After [18].

Reports in [1], [18] conclude this data suggests a common mechanism between the two devices. Structurally, the SiC Diode does not have a parasitic BJT suggesting this cannot be the mechanism responsible in the SiC MOSFET. Analytical studies in [20] showed that SiC devices experience heat generation density 100x faster than Si showing burnout on two separate timescales. Additionally, it's been shown experimentally in [1], [48], [49] that SiC devices are more prone to permanent degradation damage than Si devices. These characteristics suggest different physical mechanisms take place causing either partial destruction of the device or catastrophic burnout.

Shoji was one of the first to suggest an alternative failure mechanism for the SiC MOSFET. As mentioned above, Shoji performed a SEB simulations on diode-like MOSFETs that had removed the n+ source to eliminate the parasitic BJT and showed no increase in SEB tolerance. The traditional MOSFET and diode-like structures compared in Fig. 6 show little difference in SEB current flow and maximum lattice temperature.

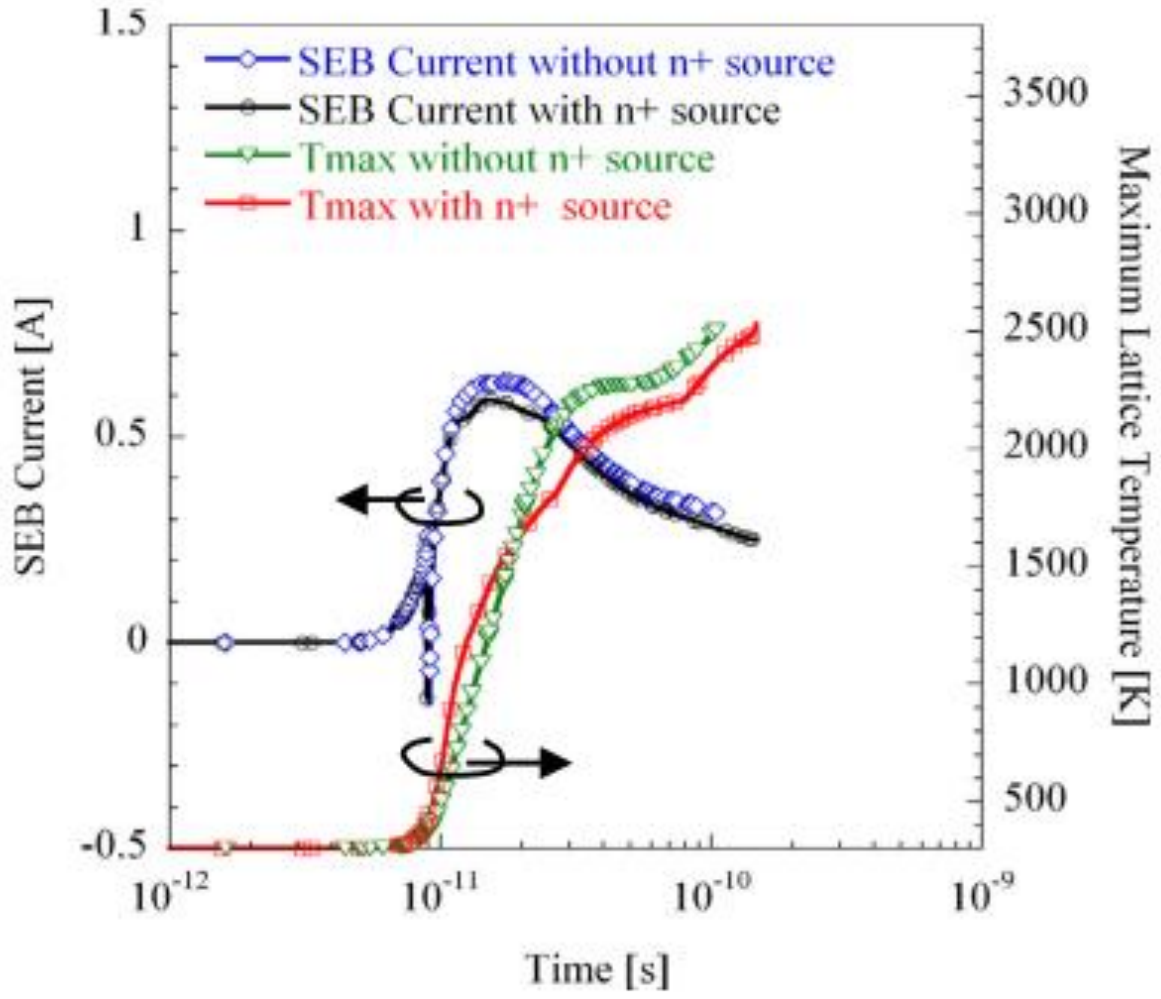

Fig. 6 Simulated SEB currents and max lattice temperatures with and without the n+ region in SiC MOSFET. After [24].

Shoji argues these results indicate little contribution from the parasitic npn-transistor towards SEB in SiC power MOSFETs. Using experimental and simulation results, Shoji observed the sublimation temperature is rapidly reached at the n/n+ homojunction. During the strike, simulation revealed that the mass injection of carriers causes a "punch through" peak electric field at this region. The spike in electric field is sufficient enough to initiate impact ionization causing high electron currents into the n/n+ interface. High current under a high electric field leads to extreme heating of the lattice until the device is thermally destructed. Shoji's high-fidelity simulation and experimental capabilities support the claims he is able to make.

In 2020, McPherson *et al.* [25] demonstrated SiC MOSFET failure mechanisms via 3D simulation that highly agree with Shoji *et al.* [20]. McPherson simulated a 1200 V SiC MOSFET under heavy ion strike of LET 46 MeV-cm$^2$/mg to correspond with experimental data collected in [50]. McPherson showed burnout at 500V and later proposes design changes to increase the SEB voltage. In the simulation from the traditional MOSFET burnout, McPherson provides time evolution 2D and 1D plots of the temperature and electric field. McPherson describes the burnout in three phases: plasma formation at initial strike injecting an excess amount of carriers, transit of carriers to their respective terminals causing modulation of the electric field and impact ionization induced by the new peak electric field at the n/n+ region leading to a regenerative thermal carrier generation process until the sublimation temperature is reached. Instead of a BJT turn on at the n+/p/n junction, McPherson describes a shorting phenomenon between the n+ source and epi region from carrier flooding that overwhelms the P-body. This shorting phenomenon diffuses carriers into the drift region from the source. McPherson observes SEB at 120ps which agrees with other simulation studies done [8], [20]. This burnout timescale is orders of magnitude too fast to be induced by a BJT turn-on.

The timescale of the BJT was also investigated by Ball *et al*. [18]. In this study, Ball attempted to improve SiC MOSFET susceptibility by implementing an in-line resistor with the source contact. This technique worked in Si as shown in [11], [39], [40] because it was able to suppress the parasitic BJT current upon activation. The data in Fig. 7, showed no change in the SEB voltage with the inline resistor and suggests that the timescale for the resistor to work is too slow (on the order of nano seconds) to assist with SiC burnout current.

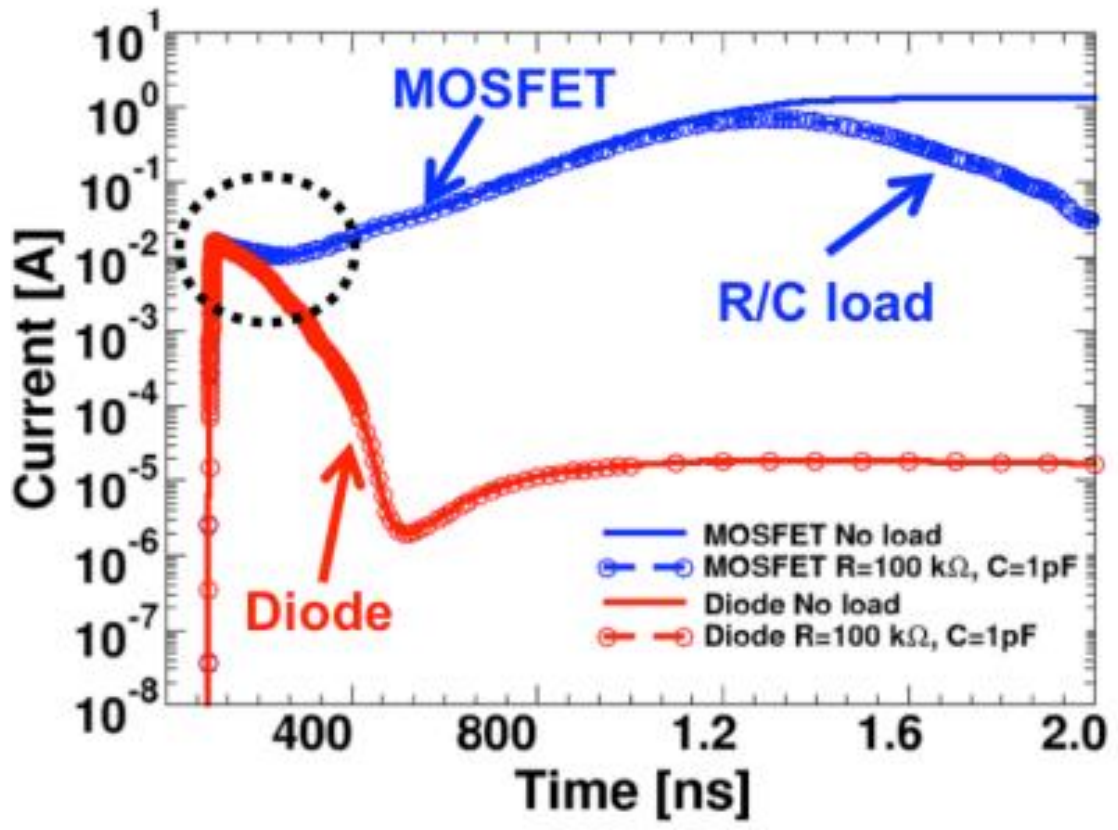

Fig. 7 Current plot comparison of Diode & MOSFET with and without implementation of the inline resistor. After [18].

Instead of the parasitic transistor, Ball proposes a high-pulse energy failure mechanism. Applying Kirchoff's rule, Ball states that the high leakage current generated during ion strike while under high voltage bias results in energy dissipation that can exceed the capabilities of the semiconductor. Ball observed high current transients that occur due to resistive shunts or conductive channels that develop on the picosecond timescale. Ball proposed the current transients during high voltage cumulate high energy pulses that eventually compromise the device. Ball created an energy dissipation SEB threshold plot shown in Fig. 8 describing the total energy dissipation required to onset degradation and SEB.

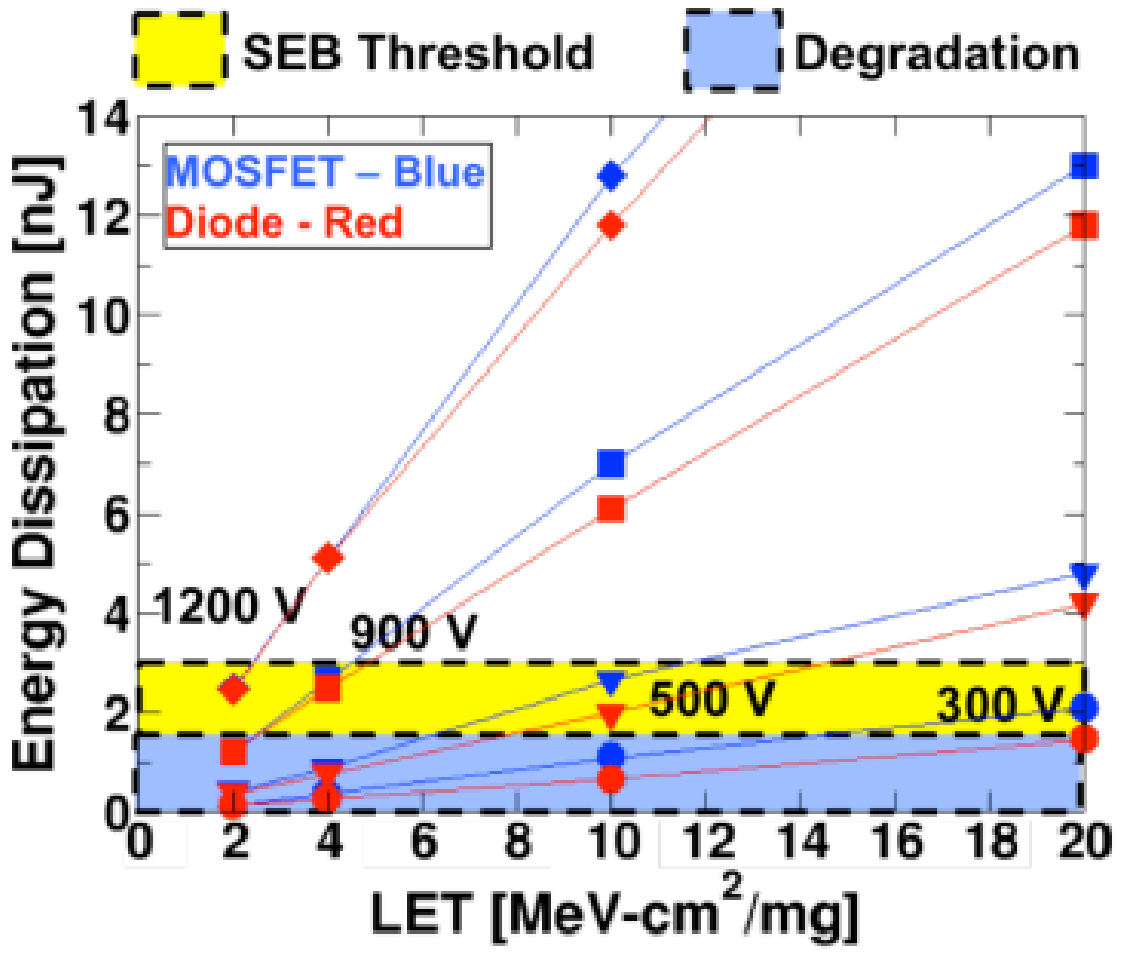

Fig. 8 Calculated Energy dissipation for 10ps after heavy ion strike with SEB threshold conditions. After [18].

The failure mechanism debate for SEB in SiC MOSFETs has gone on since the turn of the century and still continues. The conventional way of thinking has been questioned as more thorough experimental and simulation studies have been performed. Although the BJT turn-on has had support in the past and has been well documented in Si, experimental studies and high-fidelity simulations have come out to support alternative mechanisms that are similar to the physics in the Si and SiC diode. The same experiments that verified the BJT turn-on in Si MOSFET SEB have shown contradicting results when performed on SiC MOSFET. This section of the paper serves as a record of these studies and the arguments made to support and refute the proposed power device failure mechanisms.

## III. SEB Performance Review of Power Device Designs

The ideal goal of radiation-hardening is to expand the SEB threshold voltage to its rated breakdown voltage with little to no degradation of the leakage current. Additionally, the device should still have the same performance in its forward and blocking states under static conditions. Figure of merits (FOM) typically used to optimize rad-hard devices consider the SEB threshold voltage with respect to the specific on resistance ($R_{on,sp}$) [4], [51], [52], [53]. While the SEB threshold voltage quantifies the radiation tolerance, the $R_{on,sp}$ quantifies the power loss during normal operation. The higher the $R_{on,sp}$ the worse the performance. It is imperative that the evaluation of simulations employed be hyper-critical. Criteria for high-fidelity simulations and future work needed to bolster radiation-effect modeling is provided in Section IV. This section serves as a review of the SiC MOS structures and proposed hardened designs SEE performances.

### Baseline VDMOSFET SEB Performance

The effects of radiation on Si and SiC VDMOSFET have been thoroughly simulated and experimentally studied. To classify the performance of a hardened SiC DMOSFET design, the baseline SiC structure must be quantified. For a 1200 V device, experiments have shown that SEB consistently occurs around 500 to 550 V, around 40% of its rated breakdown [1], [10], [17], [18], [19]. Fig. 9 shows the thresholds that a 1200V SiC DMOSFET experience degradation and burnout with respect to heavy ion LET. The burnout threshold remains about the same for any ion LET greater than 10 MeV-cm$^2$/mg.

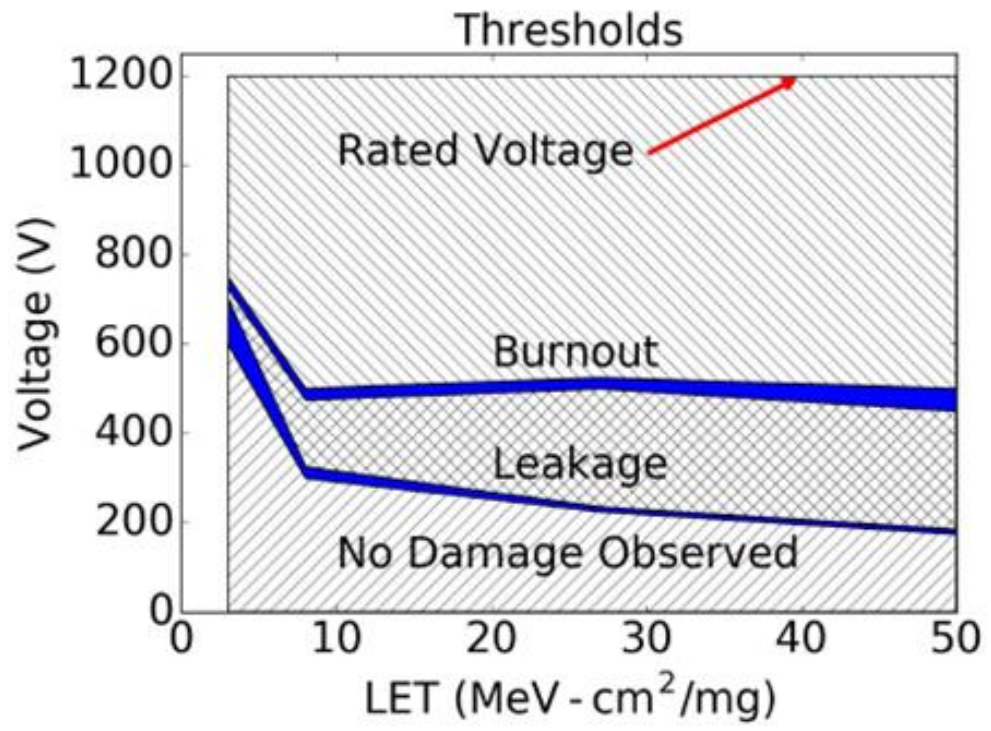

Fig. 9 SiC MOSFET Burnout Threshold. After [10].

The $R_{on,sp}$ of a commercial 1200V SiC VDMOSFET is around 2-3 mΩ-cm$^2$ [4], [54], [55], [56]. A comparison of the SiC and Si 1D performance limits was made by Gajewski *et al.* [54]. Fig. 10 shows the comparison of the $R_{on,sp}$ with respect to the breakdown voltages after Gajewski. A $10^2$ magnitude improvement is illustrated between the materials.

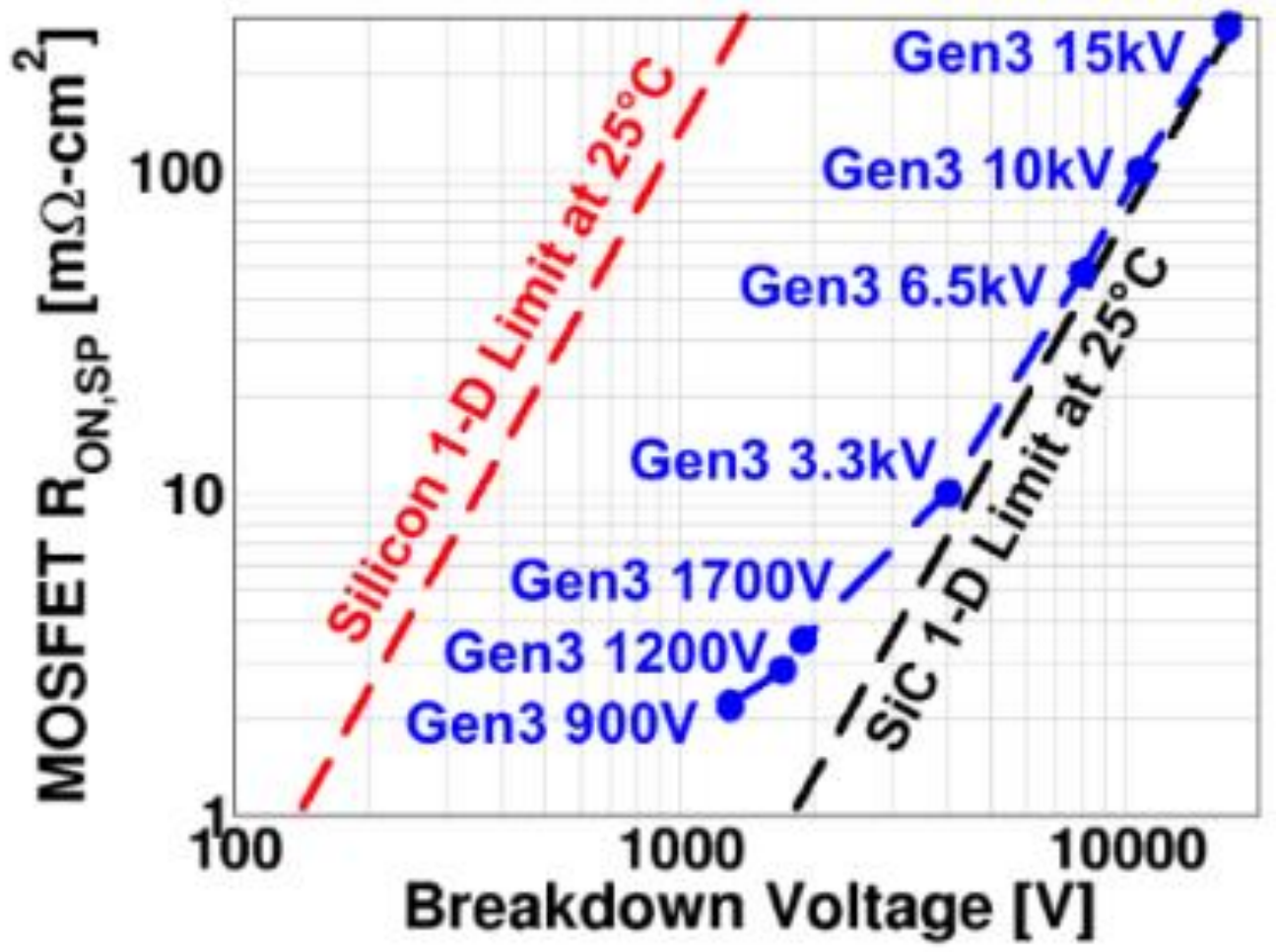

Fig. 10 1D specific on-resistance versus breakdown voltage comparing Si and SiC. After [54].

Ball *et al.* [56]. proposed using high voltage rated DMOSFETs to have a safe operating area (SOA) around the industry desired 1200V mark. The study introduced a 3300 V rated SiC MOSFET and experimentally showed the SEB threshold voltage can be improved under heavy ion strike. Although the 1200V mark was not reached, Ball showed SEB occurred at 825V and proposed a SOA of 650V (57% improvement) in intense radiation environments (LET > 10 MeV-cm$^2$/mg). However, this improvement in SOA comes with a tradeoff because a device with this breakdown has a much thicker epitaxial layer. Subsequently, the $R_{on,sp}$ of the 3300V device is 10 mΩ-cm$^2$ (compared to the 1200V 3 mΩ-cm$^2$). As a result this device exhibits 3x the resistive losses to its 1200V counterpart, making the operation of this device very impractical. This derating method was also applied via simulation in [4]. In this work, McPherson increased the thickness of the epitaxial layer on a DMOSFET until the electric field was net zero at the n/n+ interface. As a result, the breakdown voltage was 3450 V and the device experienced SEB at 1200V, while the $R_{on,sp}$ was 7.6 mΩ-cm$^2$. Recent experimental and simulation studies suggest a unified degradation threshold that exists among high-voltage power devices. Johnson *et al.* [57] found that regardless of the rated breakdown voltage of the device, undesirable leakage currents occur at 180V in intense radiation environments. Derating of the traditional DMOSFET may not be a practical method of radiation hardening due to undesirable on-state resistance losses and cumulative leakage degradation [4], [17], [57], [58].

**Hardened VDMOSFET Designs**

The definition of a hardened MOSFET is a design that has been intentionally modified to improve radiation resistance. Many techniques have been employed to correct radiation induced malfunctions that occur on power modules, however a hardened device design has not yet been employed [3], [59]. As we know in Si, the SEB performance of a MOSFETs can be improved by using an optimized buffer layer design [35], [44]. The buffer layer is typically placed between the n/n+ homojunction to reduce the peak electric field that has been shown to generate at this location and can achieve a higher second breakdown voltage. This type of design has also been simulated in SiC. Zhou *et al.* [24] looked to improve the SOA of 600V breakdown SiC DMOSFETs and was one of the first to implement a buffer layer in SiC. Using a 2D simulation, Zhou showed the SOA of a SiC MOSFET when exposed to high LET ions (.1 to 1 pC/μm) was 15V, where as the 600V Si counterpart had an SOA of 110V. These results do not align with experimental Si and SiC parts that have been tested [1], [17], [56]. To improve the SiC SOA threshold to 600V, Zhou introduced a 350 μm buffer layer doped on the order of $10^{18}$ $cm^{-3}$. Although SEB improvement was shown, the thickness of the hardened device increased the $R_{on,sp}$ to an undesirable >25%. A buffer layer that large is extremely impractical for device design and the simulation conducted is highly inaccurate. Zhou concludes alternative solutions should be implemented to consider SEB ruggedness and overall performance.

The buffer layer was also attempted by McPherson *et al.* [25]. Using high-fidelity 3D simulation, two styles of buffer layer were implemented into a 1200V rated SiC DMOSFET. One uniformly doped and one non-uniformly doped buffer layers both 2μm thick were applied at the epi/substrate interface. The baseline 1200V MOSFET struck by a 46 MeV-$cm^2$/mg ion showed burnout at 525V, while the two buffer layer designs showed improved thresholds at 900V and 925V, respectively. Both designs behaved similarly in dynamic simulation, however McPherson recommended the non-uniform buffer layer because of the less the sharp electric field peaks it generated in comparison to the baseline and uniform designs. The addition of either type of buffer layer only increased $R_{on,sp}$ by 5%. McPherson *et al.* [26] later proposed the SEB failure mechanisms for the buffer layer design. Studying a uniformly doped buffer layer design, 2μm thick doped at $4*10^{16}$ $cm^{-3}$ ($R_{on,sp}$ of 3.3mΩ-$cm^2$), McPherson observes the SEB location shift from the n/n+ region to 3-4μm from the surface in the epi layer. Impact ionization is argued to not drive the burnout here, instead it is driven by the thermal generation of carriers. McPherson describes the formation of a mesoplasma at the failure location that generates thermal carriers and increases current density to induce a thermal runaway. McPherson's most optimized buffer layer design was displayed in [4]. The 3D model had a static breakdown of 2038V, experienced SEB at 1200V and had a $R_{on,sp}$ of 3.2mΩ-$cm^2$. The buffer layer design exhibits SEB over twice as long after ion strike as the baseline design and shows strong suitability for rad-hard applications.

Additional buffer layer designs showed improved SEB tolerance in SiC MOSFET designs. Some research works have proposed improving power device's performances by introducing multiple buffer layer and dosage combinations [60], [61]. Through 2D simulation, Lu *et al.* [62] implemented three different buffer layer deigns: a single layer, double layer and triple layer, where each layer is a different doping concentration. The exact dimensions of the optimal design were not disclosed, however the buffer thickness was anywhere from 4 to 50μm and the doping concentrations ranged from $10^{16}$ to $10^{19}$ $cm^{-3}$. The baseline 1200V SiC DMOSFET showed SEB from 350-500V, depending on the ion LET. The lower threshold voltages could be owed to the 1500 K lattice temperature failure criteria Lu employed taken from [63]. The introduction of just the single buffer layer increased the SEB SOA to its 1200V rating. The other double and triple layer designs showed even better tolerance. The thickness and doping concentration added to the structure inherently increase the static breakdown voltage of the device. While the device is rated at 1200V, the breakdown voltage is much higher. Lu acknowledges the increase in $R_{on,sp}$ this design will have, but does not quantify it in this work. Intuitively, a device this thick and with that many doping regions will have a considerably undesirable electrical performance. Disclosure of the optimized design parameters is needed to accurately assess this designs suitability for rad-hard applications.

2D simulations where isothermal effects are assumed have shown very tolerable SEB designs [7], [64]. In a 2.5D and isothermal simulation, Zhu *et al.* [7] demonstrated a channel narrowing technique in 1200V SiC DMOSFETs that improves the SEE threshold from 500/600V to >1200V. This type of technique can modulate the electric field within the device to improve SEB tolerance, but trades off with the $R_{on,sp}$ of the device [7], [65]. Although the baseline design aligns with experimental data, the ion strike in this simulation environment causes the electro-thermal response to be non-physical. In turn, the material properties are temperature independent and the lattice temperature of the device is under-assumed. This is problematic when accurately trying to determine SEB. In order to accurately assess the effect of this design technique, a 2D axisymmetric or 3D simulation with temperature dependence must be employed.

Research into localized carrier lifetime control suggested the Si VDMOSFETs can improve the SEB threshold and SOA [66]. McPherson *et al.* [26], investigated SiC material properties effect on SEB by artificially fixing them via 3D simulation. A 100x increase or decrease of the carrier lifetime had a minimal effect (25V decrease and 25V increase, respectively) on the relative SEB threshold voltage. The most profound effect on SEB threshold voltage was seen by a 10x decrease in saturation velocity or a fixed 4.9 W/cm-K thermal conductivity value in the SiC lattice. The change in either of these two parameters yielded an additional +650V and +700V, respectively, to the baseline SEB threshold voltage. Despite the improvement to threshold voltage by decreasing the saturation velocity, there exists a trend for semiconductor materials where the saturation velocity increases as the bandgap increases. Since WBG materials have the best electronic performances, wider band-gap semiconductors are always going to be preferred in power device applications so this saturation velocity will never be able to decrease to the level performed in this simulation [67]. However, thermal conductivity is a more plausible route as there are many semiconductor materials with extremely resilient thermal conductivities. Diamond, for example, has been shown to be one of the best semiconductor materials known on the planet and could potentially solve the rad-hard dilemma if the jewelry market was not so lucrative [68].

**Baseline Trench Gate MOSFET SEB Performance**

The trench gate or UMOSFET has been looked at as an alternative to the DMOSFET in power electronics because it can reduce the $R_{on,sp}$ owing to its smaller cell pitch and high channel mobility on the trench sidewall [64], [65], [69],. Experimental studies on Si UMOSFETs show their susceptibility to degradation and catastrophic SEEs under radiation [70], [71], [72], [73], [74], [75], [76], [77] and revealed similar performances to the VDMOSFET. A good review of the Si UMOS radiation failure and degradation is outlined in [71]. Experimental SEE studies on SiC MOS structures primarily focus on the planar structure leaving little available on the trench gate. Preliminary results on the SiC UMOS are studied via simulation [65]. Experimental and simulation studies on the UMOSFET show greater susceptibility to single event gate rupture (SEGR) [65], [71]. The high electric field in the gate oxide of the SiC trench MOSFET is a major concern when operating under reverse blocking mode [65], [78]. Zhou *et al.* [65] simulated 650V SiC Trench MOSFETs under heavy ion strike (LET of 1pC/μm) that showed a safe operating area at 70V (11% of $V_{BR}$) for SEB and 20V (3% of $V_{BR}$) for SEGR. Kim *et al.* [83] via 2D simulation showed 1200V rated SiC UMOSFETs have a SEB threshold of 530V, matching that of its planar counterpart.

Wang *et al.* [79] studying the same device showed a SEB threshold voltage of 450V. However, Wang's model showed a max lattice temperature of 492K at SEB, well below the sublimation temperature of SiC, suggesting a disagreement with known planar MOSFET failure mechanisms.

**Hardened Trench Gate MOSFET Designs**

There are many hardening techniques that have been employed on Si UMOSFETs to improve SEE tolerance, however a review of these designs is not within the scope of this paper and are more properly discussed in [71]. In SiC, few studies have been conducted. Owing to the high electric field, improvement of the catastrophic SEE threshold is particularly difficult in the UMOSFET. Wang *et al.* [80] showed baseline SiC UMOSFET designs with static breakdown of 1217V showed that by 300V (LET of .5pC/μm) the sublimation temperature in the lattice had been far surpassed and damaging electric field peaks had been reached in the oxide. The $R_{on,sp}$ of the baseline design was 10mΩ-cm$^2$, more than three times the typical UMOSFET and DMOSFET designs induce [64], [65], [69]. The improved design proposed a multi-buffer layer and P+ shielding region that wraps the bottom of the trench gate. The methodology behind the buffer layers is the same as the DMOSFET. The P+ shielding region was added to suppress the electric field at the oxide. The improved design had a $R_{on,sp}$ of 6 mΩ-cm$^2$, static breakdown of 1257V and SEB threshold of 650V. It was noted an uncontrollable high regenerative current triggered by a BJT turn on was not observed. Instead, the device was thermally damaged before the parasitic transistor was even fully turned on. Although there is considerable improvement to the SEB threshold it is still not at industry's desire. Additionally, the $R_{on,sp}$ is impractical for general usage and more tests should be done to consider the effects of SEGR on this particular device. Wang concludes the 2D simulations conducted are a qualitative characterization of a heavy ion strike.

Previously, Wang *et al.* [79] had proposed island buffer layers in the trench-gate MOSFET. In this 2D simulation work, the initial designs with a breakdown of 1200V experienced SEB at 450V. Although Wang deduced SEB by tracking drain current, the maximum global device temperature only reached 492K within the device at burnout. The temperature dependence of the models employed was not disclosed. Initial improved designs in this study added two buffer layers each 1μm thick doped at 5e16cm$^{-3}$ and 3e17cm$^{-3}$ brought the SEB threshold to 550V. The most successful SEB design was the island buffer layer displayed in Fig. 11. This design incorporates a heavily doped N+ region within a lighter doped N buffer region, which drastically reduces electric field peaks during ion strike. The most optimized design ($V_{BR}$ 1183V), parameters disclosed in Fig. 11, improved the SEB threshold voltage to 660V (47% improvement).

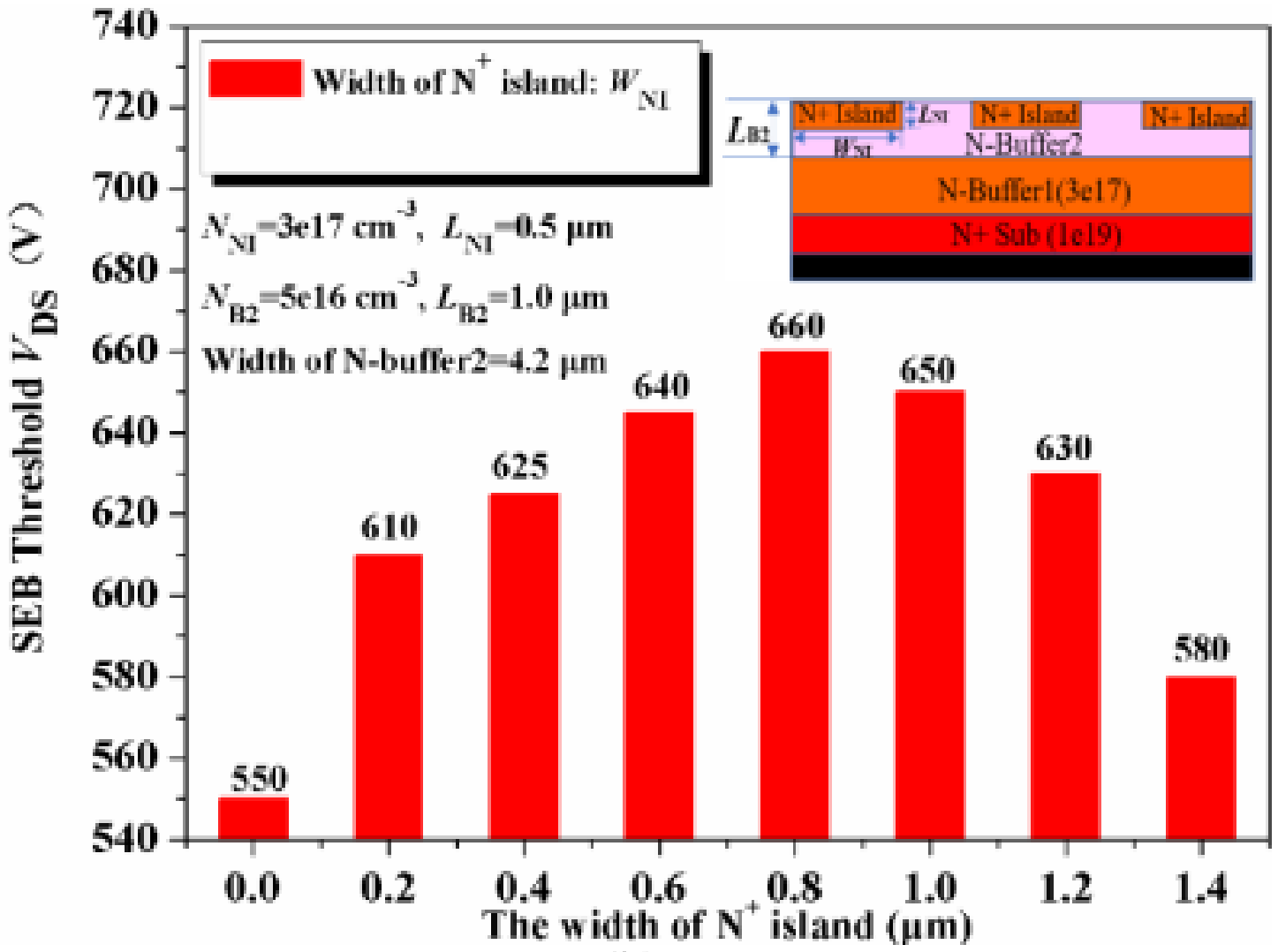

Fig. 11 Island Buffer Layer Results and Optimized Design Parameters. After [79].

The baseline design had an $R_{on,sp}$ of 2.91 mΩ-cm$^2$; the island designs $R_{on,sp}$ was higher but was not quantified. This congested buffer layer scheme more than likely increases $R_{on,sp}$ to an undesirable amount. Additionally, Wang's max temperature at SEB was on the order of hundreds leaving skepticism to the simulation parameters employed. More works have implemented buffer layer designs to see SEB improvement in the SiC UMOSFET [65], [81].

Kim *et al.* [82] implemented a self-aligned sidewall heterojunction diode due to its excellent body diode characteristics and its reduction of bipolar degradation effect from work shown in [83]. Kim saw a 25% increase to the SEB threshold as compared to the traditional double trench MOSFET with this design. However, the $R_{on,sp}$ in Kim's baseline and improved designs were both unacceptably at 16.55 mΩ-cm$^2$.

Another method of SEE hardening proposed by Zhou *et al.* [65] suggested adjusting the mesa width ($L_{mesa}$). The mesa width is the length between the P+ and the gate underneath the P-base and source contact in a double trench MOSFET. The $L_{mesa}$ plays a critical role in modulating the peak electric field in the gate oxide and can be changed to help improve SEE ruggedness in a design. Zhou showed by decreasing the width at this interface, the SOA for SEB in a 650V design was improved from 70V to 300V with a 56% increase in $R_{on,sp}$. The SOA for SEGR with the smaller $L_{mesa}$ only improved from 20V to 35V. Although such an increase in $R_{on,sp}$ would still be better than the Si UMOSFET counterpart, the SOA for this device renders it impractical. Overall, preliminary SiC simulations suggest the electric field at the gate oxide in a trench gate MOSFET makes this device far too susceptible to SEEs. For high-power applications in intense radiation environments, alternative MOSFET structures should be preferred. More experimental studies on SiC UMOSFETs are needed to verify the simulation work mentioned above and to push this device structure forward.

**Baseline Super Junction MOSFET SEB Performance**

It has been shown in the above sections that an improved SEB tolerance can be achieved with the traditional MOSFETs by increasing the breakdown voltage. However, to reach the 1200V SEB rating with these methods, the $R_{on,sp}$ increases over 100% [4]. A potential solution to this tradeoff has been proposed with a super junction (SJ) structure. The SJ structure replaces the uniformly doped epitaxial layer with an array of alternating N+ and P+ pillars. This configuration has horizontal and vertical electric fields giving the device a rectangular electric field profile. Consequently, the doping concentrations of the pillars can be higher than conventional epitaxial layers yielding a significant reduction in $R_{on,sp}$ without effecting the breakdown voltage. The SJ MOSFET offers reduced gate and output charges which allows for more efficient switching at higher and lower frequencies [84]. The first demonstration of this structure in Si was in the late 1990s [85]. Due to the unique electric field profile and space charge balance in the device, Huang *et al.* [86] hypothesized the Si SJ MOSFET would have a greater SEB tolerance than the VDMOS and concluded this was so after 2D heavy ion strike simulation. Ikeda *et al.* [87] later showed with experimental results the SJ had no improvement to SEB tolerance over the VDMOS. Further simulation and experimental studies on the Si SJ MOSFET showed there was no inherent advantage to SEE tolerance [88], [89], [90], [91], [92]. More recently, it's been demonstrated SJ theory can be effective for SiC Power MOSFETs [93] and the first SiC SJ MOSFET was demonstrated in [94] at the 2016 ESCRM. Zhou *et al.* [95] showed the reduction in $R_{on,sp}$ for the SiC SJ MOSFET can actually surpass the 1D limit ($R_{1DLimit}$) in SiC and is illustrated in Fig. 12.

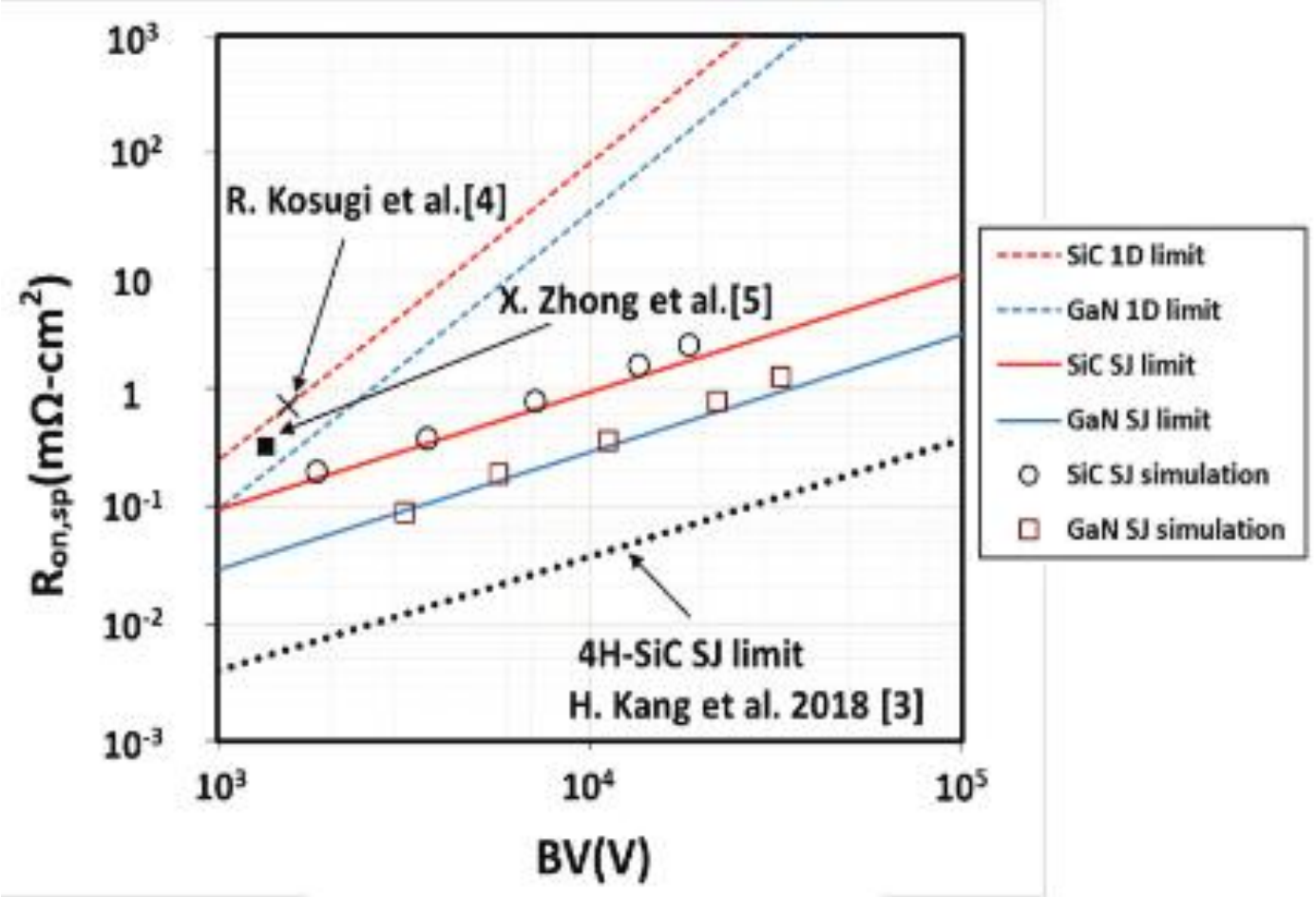

Fig 12. WBG 1D limit and SJ structure comparison. After [95].

This allows for the derating method that showed SEB tolerance improvement for the traditional MOSFET to be implemented without the costly $R_{on,sp}$ tradeoff. McPherson *et al.* [51] demonstrated a SiC SJ DMOSFET 3D design that had a breakdown voltage of 5932V, experienced SEB (LET 46 MeV-cm$^2$/mg) at 1150V, while the $R_{on,sp}$ was only 3.0 mΩ-cm$^2$. In this work McPherson, found a relationship between pillar width, SEB threshold voltage ($V_{SEB}$) and $R_{on,sp}$ and maximized these values to achieve the best design by implementing a FOM that takes the ($V_{SEB}$/BV) with respect to the ($R_{1DLimit}/R_{on,sp}$) for each design iteration. This FOM is quite useful for selecting suitable rad-hard designs and is suggested for future work on this technology. The SEB failure mechanism described in this device is very similar to the failure within the SiC diode. Here, a mesoplasma forms from simultaneous high current and high electric field at the pillar substrate interface leading to a thermal runaway until the sublimation point is reached. Other SiC SJ MOSFET simulation work have also described the SEB failure mechanism from a critical temperature standpoint [53]. Experimental and simulation studies suggest SEGR is the most prone SEE in the SJ MOSFET and could cause earlier burnout than SEB simulations test for [89]. Due to the novelty of the design, the fabrication of a SiC SJ MOSFET has not been commercialized and experimental SEE testing has not been pursued, however is needed for validation of the SEB success in these devices.

**Hardened Super Junction MOSFET Design**

Most hardened designs for the SJ structure have been studied on Si. Muthuseenu *et al*. [92] showed SEE improvements in Si from the adoption of a trench-gate, buffer layer and modified P-body combination. This change in $R_{on,sp}$ was not quantified for this modified design and the ion strikes tested were in a isothermal 2D simulation. Although the study showed improvement from the design, fidelity in the simulation cannot be attributed. Wang *et al*. [88] implemented two different buffer layer designs on Si SJ MOSFETs of 32 and 45µm thick of uniform and graded doping, respectively, to investigate its effect on $V_{SEB}$ via 2D simulation. The resulting structure had a >25% increase in $R_{on,sp}$, however increased the SEB threshold for the 650V rated device from 103V to 605V and 695V for the two designs, respectively. Wang also saw the same improvements for a novel P-type buried layer design. Its effect on $R_{on,sp}$ is not known and the thermal assumptions in the simulation were not disclosed, however are needed for a fair comprehensive evaluation.

Few studies have been employed to investigate the SEB performance of the SiC SJ MOSFET. No hardened structures were suggested by McPherson in [51]. In later works, McPherson *et al.* [96] proposed a semi super junction design that was tested via 3D simulation. Similar to a buffer layer, the semi SJ design inserts a uniformly doped layer between the SJ pillars and the substrate. The structure employed is shown in Fig. 13.

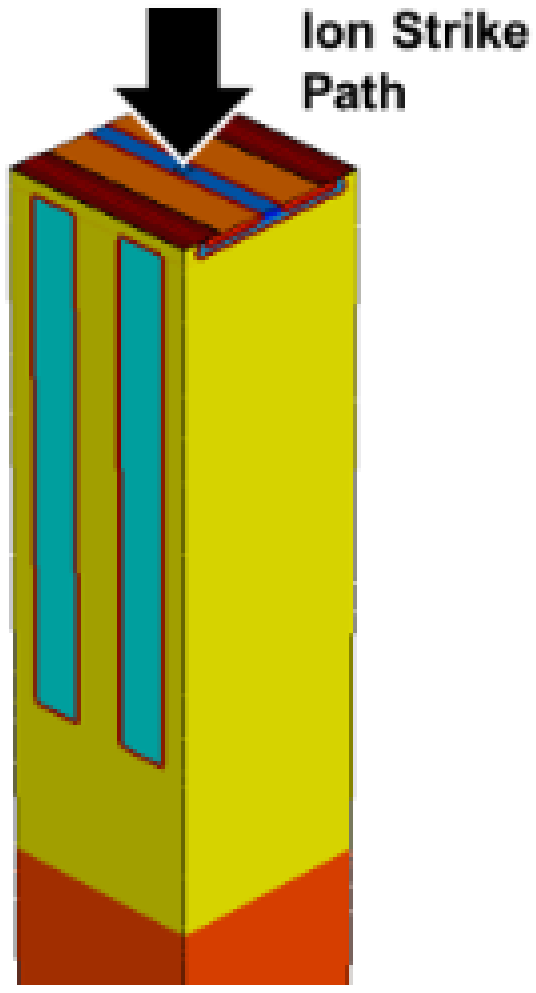

Fig. 13 3D view of Semi SJ model. After [96].

The most optimized semi SJ model had alternating 1.2µm pillars cover 90% of the drift layer, leaving the bottom 10% uniformly N doped. The baseline SJ model had a breakdown of 5077V, $V_{SEB}$ of 950V and $R_{on,sp}$ of 3.2 mΩ-cm$^2$. The most optimal semi SJ design had a BV of 4897V, $V_{SEB}$ of 1750V and $R_{on,sp}$ of 3.2 mΩ-cm$^2$. For the same on-state performance this SJ design showed an 84% increase in $V_{SEB}$, leaving the SOA of this device near the desire of industry. It was noted a DMOSFET with a drift layer as thick as the semi SJ design (31µm) has a BV of 3354V, $V_{SEB}$ of 1250V and $R_{on,sp}$ of 8.0 mΩ-cm$^2$. This comparison highlights the overt benefit and suitability this semi SJ design provides. More recently, Yu *et al.* [53] implemented a buffer layer on a SiC SJ MOSFET to see SEB improvement. In 2D simulation, a 5µm uniformly doped buffer layer improved degeneration tolerance from 500V to 800V on a 1200V rated design. Yu observed failure at the SiC/metal contact interface (>1500 K) and advised this be considered in future designs. The SJ design has a major $R_{1DLimit}$ to $R_{on,sp}$ advantage over traditional device designs, suggesting high voltage SJ and hardened SJ designs are promising candidates for radiation hardening applications.

## IV. High Fidelity Simulations

Modeling the proton, neutron and heavy ion transport process to analyze the SEBs in SiC devices can be done with simulation technology and software. Inputting known energy distributions, these TCAD programs have the ability to perform coupled transient simulations of high-energy particle strikes on semiconductor power devices. Modeling the SiC MOSFET during high energy particle impact with a simulator can provide insight to the failure mechanisms responsible for device heating and catastrophic failure [87], [97]. Using that insight, design configurations can be implemented to help mitigate these effects and improve SEB tolerance. However, there is still a necessary process that needs to be completed to assure high-fidelity within a simulation. The use of 2D simulation has raised skepticism by researchers because the deposition of heavy ion strikes are inherently three dimensional in nature [4], [29]. In a 2D simulation, the particle strike is read as a plane versus as a point charge in 3D. Thus, the transport processes in 2D simulations have difficulty accurately estimating the effects of corrupting plasma and device electric field. It is highly suggested that 2D results be taken qualitatively instead of quantitatively [78], [80], [81], [88]. To verify a simulation, the parameters and set-up need to align with an experimental study previously done. The devices design and rated breakdown voltage need to be identical and verified separately in a static simulation. In the dynamic simulation, the radiation type, energy, LET and SEB threshold need to match those done in the experimental study. Once this step has been taken, fidelity can be attributed to the simulation and design tactics can be implemented to help improve SEB tolerance. It should also be noted that parameters, such as, but not limited to, impact ionization coefficients, only have data available up to relatively low temperatures. Although there are default values and models available to replicate the device physics under dynamic conditions it is only an extrapolation of the known data. This should be considered with all simulations when assessing the suitability and merit of design results from simulations.

The ranges of devices and materials in use or envisioned for use, coupled with the complexity of radiation damage effects on the properties and performance, severely limit what can be learned from just experiments. Existing tools for modeling and simulation must be adapted to study radiation damage with an atomistic perspective. Due to recent technological advances in the past five years, it is not hard to believe more advances can be accomplished. Ideal test structures can now be produced from the progress that’s been made in synthesis and fabrication that allow precise control of chemistry and structure. Individual point defects caused by radiation events can now be imaged from new scanning and transmission electron microscopy techniques. New atomic scale modeling can make use of these new experimentally developed methods to give a comprehensive carrier capture and inelastic-scattering cross section theory. It is now feasible to develop device-scale Monte Carlo codes that incorporate full band structures. The various new inputs can be used in multiscale engineering-level models for device degradation. Development of this technology will ultimately lead to the highest precision SEE simulations that will inevitably clarify discrepancies in device failure mechanisms and lead to designs of reliable and radiation tolerant power devices.

**Failure Criteria**

In dynamic simulation, a failure criterion needs to be identified to signal the point at which SEB occurs. The use of SiC sublimation temperature [4], [8], [20], [25], [26] critical electric field [18] and leakage current [21], [22], [23], [24] have all been previously used. The parameter critical electric field is a derived value based on the devices design in static conditions. As shown in Eq. 3, the analytical critical electric field ($E_c$) is a function of the depletion width at breakdown ($W_{BV}$).

$$E_c = 1.86 * 10^{-7}\ W_{BV}^{1/6} \tag{3}$$

During a dynamic event, like a heavy ion strike, the electric field can reach extreme levels far surpassing the static critical electric field without triggering an SEB because of the narrowing of epitaxial layer during impact ionization. This narrowing can peak the electric field to see levels higher than the static critical electric field, but will not trigger an avalanche event. This makes this parameter highly unreliable in determining when SEB has terminated a power device.

Monitoring the drain current for runaway is another common method of determining an SEB event. When the drain current increases to a high enough magnitude while the device in its blocking state, energy begins to dissipate that eventually leads to the thermal destruction of the device. The equations in the simulator that output these values are dependent upon the temperature, power dissipation and thermal properties of the device. This is an accurate method of SEB measurement and can give an idea of the SEB timescale, but is highly dependent on more potent characteristics and cannot detail the exact moment SEB occurs.

As shown experimentally in [1], [8], [20] SEB is inherently a thermally terminated event. In post tear down of the device, the SiC MOSFET has shown cracks as a result of expansion stress due to the sublimation of SiC. Stereomicroscopy and slice-and-view techniques revealed, molten SiC has forms in the drift region from the rapid temperature increase within the device. Simulations using the SiC sublimation temperature as the SEB failure criteria, which details the exact moments and timescale of when SEB occurs. This is amongst the most accurate methods of SEB determination. Monitoring all three of these characteristics in conjunction is suggested for proper SEB analysis.

## V. Conclusion

This review summarizes the current state of knowledge regarding the failure mechanisms of silicon carbide power devices from heavy-ion exposure and hardened designs proposed to improve radiation tolerance. Published work has demonstrated MOSFET vulnerability to SEEs and catastrophic failures from SEB and SEGR. The physics of SEB in SiC have shown different characteristics than in Si and has led to debate regarding the source of the regenerative current leading to device burnout. Although it was originally thought the BJT turn-on was the primary mechanism for SEB, recent high-fidelity simulation studies strongly support an alternative mechanism caused by the simultaneous high current and high electric field yielding energy and temperatures surpassing the limit of the material. Although there is no current way to experimentally verify the exact mechanism, advances in simulation capabilities in the future will be able to provide more fidelity to this understanding.

Many designs have been employed to improve the MOSFETs SEB and SEGR threshold, however few have suitable on-state performance for device operation under static conditions. Promising designs include, first, a VDMOSFET with a thin buffer layer that adds <5% to the $R_{on,sp}$. Fabrication of this design is manageable and should be the next step in validating the devices performance. Additionally, the SiC SJ and hardened SJ MOSFETs have shown very promising results providing high SEB threshold with no sacrifice in $R_{on,sp}$. However, fabrication of the SJ designs may be complex and may not be cost effective. Conclusive support of rad-hard designs can be drawn once experimental static and SEB tests are conducted. Additionally, more work must be done to understand the tolerance to onset degradation that occurs before catastrophic events that effects both hardened and non-hardened device performance. Additional refinement of the models for simulation and experimental methods will advance SiC MOSFET hardened design analysis and development. This review stands as an objective evaluation of this technology in order to help provide better selection criteria for radiation-bound power systems.

**Acknowledgement**

The authors would like to thank the colleagues at Rensselaer Polytechnic Institute, in particular Dr. Joseph Alexander McPherson (from NuCoMP lab) and Dr. Paul Chow (from EECS Department), for helpful discussions. The presented work was conducted during the graduate study of the first author at Rensselaer Polytechnic Institute. The first author was supported by the United States Nuclear Regulatory Commission Fellowship Program under Grant Number 31310018M0003.

## References

[1] J.-M. Lauenstein *et al.*, "Wide-bandgap-power-SiC and GaN-radiation reliability," *IEEE Nucleaer Space Radiation Effects Short Course,* 2020.

[2] J. Y. Tsao *et al.*, "Ultrawide-Bandgap Semiconductors: Research Opportunities and Challenges," *Advanced Electronic Materials,* vol. 4, no. 1, 2018.

[3] NASA, "Technology Roadmaps—TA 3: Space Power and Energy Storage," NASA Technology Roadmaps, 2015.

[4] J. A. McPherson, "Modeling and Improving Single-Event Burnout Performance from Heavy Ion Bombardment in High-Voltage 4h-SiC Power Devices," PhD Dissertation, Rensselaer Polytechnic Institute, 2021.

[5] T. Nitta *et al.*, "Cosmic Ray Failure Mechanism and Critical Factors for 3.3kV Hybrid SiC Modules," in *PCIM Europe*, 2016.

[6] A. F. Witulski *et al.*, "Single-Event Burnout Mechanisms in SiC Power MOSFETs," *IEEE Transactions on Nuclear Scienceq,* vol. 65, no. 8, pp. 1951-1955, 2018.

[7] X. Zhu *et al.*, "Radiation hardness study on SiC power MOSFETs," in *International Conference on Silicon Carbide and Related Materials*, 2018.

[8] T. Shoji *et al.*, "Experimental and simulation studies of neutron-induced single-event burnout in SiC power diodes," *Japanese Journal of Applied Physics,* vol. 53, no. 4S, p. 04EP03, 2014.

[9] D. J. Lichtenwater *et al.*, "Reliability of SiC Power Devices against Cosmic Ray Neutron Single-Event Burnout," *Material Science Forum,* vol. 924, pp. 559-562, 2018.

[10] J.-M. Lauenstein *et al.*, "Taking SiC power devices to the final frontier: Addressing challenges of the space radiation environment," No. GSFC-E-DAA-TN47451, 2017.

[11] J. L. Titus *et al.*, "Single-event burnout of power bipolar junction transistors," *IEEE transactions on nuclear science,* vol. 38, no. 6, pp. 1315-1322, 1991.

[12] G. H. Johnson, "Features of a heavy-ion-generated-current filament used in modeling single-event burnout of power MOSFETs," PhD Dissertation, University of Arizona, 1990.

[13] J. H. Hohl *et al.*, "Features of the triggering mechanism for single event burnout of power MOSFETs," *IEEE Transactions on Nuclear Science,* vol. 36, no. 6, pp. 2260-2266, 1989.

[14] J. W. Waskiewicz *et al.*, "Burnout of Power MOS Transistors with Heavy Ions of 252-Cf," *IEEE Transactions on Nuclear Science,* vol. 33, no. 6, pp. 1710-1713, 1986.

[15] S. Kuboyama *et al.*, "Mechanism for single-event burnout of power MOSFETs and its characterization technique," *IEEE Transactions on Nuclear Science,* vol. 39, no. 6, pp. 1698-1703, 1993.

[16] S. Kuboyama *et al.*, "Numerical analysis of single event burnout of power MOSFETs," *IEEE Transactions on Nuclear Science,* vol. 40, no. 6, pp. 1872-1879, 1993.

[17] M. C. Casey *et al.*, "Schottky Diode Derating for Survivability in a Heavy-Ion Environment," *IEEE Transactions on Nuclear Science,* vol. 62, no. 6, pp. 2482-2489, 2015.

[18] D. R. Ball *et al.*, "Ion-Induced Energy Pulse Mechanism for Single-Event Burnout in High-Voltage SiC Power MOSFETs and Junction Barrier Schottky Diodes," *IEEE Transactions on Nuclear Science,* vol. 67, no. 1, pp. 22-28, 2020.

[19] J.-M. Lauenstein *et al.*, "Single-Event Effects in Silicon and Silicon Carbide Power Devices," in *NEPP Electronic Technology Workshop,* 2014.

[20] T. Shoji *et al.*, "Analysis of neutron induced single-event burnout in SiC power MOSFETs," *Microelectronics Reliability,* vol. 55, no. 9-10, pp. 1517-1521, 2015.

[21] X. Zhang, "Failure mechanism investigation for silicon carbide power devices," MS Dissertation, University of Maryland, 2006.

[22] A. Griffoni *et al.*, "Neutron-induced failure in super-junction, IGBT, and SiC power devices," in *12th European Conference on Radiation and Its Effects on Components and Systems,* 2011.

[23] R. A. Johnson *et al.*, "Enhanced charge collection in SiC power MOSFETs demonstrated by pulse-laser two-photon absorption SEE experiments," *IEEE Transactions on Nuclear Science,* vol. 66, no. 7, pp. 1694-1701, 2019.

[24] X. Zhou *et al.*, "A simulation-based comparison between Si and SiC MOSFETs on single-event burnout susceptibility," *IEEE Transactions on Electron Devices,* vol. 66, no. 6, pp. 2551-2556, 2019.

[25] J. A. McPherson *et al.*, "Mechanisms of Heavy Ion-Induced Single Event Burnout in 4H-SiC Power MOSFETs," *Material Science Forum,* vol. 1004, pp. 889-896, 2020.

[26] J. A. McPherson *et al.*, "Ion-Induced Mesoplasma Formation and Thermal Destruction in 4HSiC Power MOSFET Devices," *IEEE Transactions on Nuclear Science,* vol. 68, no. 5, pp. 651-658, 2021.

[27] M. Pocaterra *et al.*, "Single event burnout failures caused in silicon carbide power devices by alpha particles emitted from radionuclides," *e-Prime-Advances in Electrical Engineering, Electronics and Energy,* p. 100203, 2023.

[28] H. Zhang *et al.*, "Experiment and simulation on degradation and burnout mechanisms of SiC MOSFET under heavy ion irradiation," *Chinese Physics B,* vol. 32, no. 2, 2023.

[29] H. Kabza *et al.*, "Cosmic radiation as a cause for power device failure and possible counter measures," in *Proceedings of the 6th International Symposium on Power Semiconductor Devices and Ics,* 1994.

[30] S. Kuboyama, *et al.* "Improved model for single-event burnout mechanism," *IEEE Transactions on Nuclear Science,* vol. 51, no. 6, pp. 3336-3341, 2004.

[31] J. L. Titus *et al.*, "Development of cosmic ray hardened power MOSFETs," *IEEE Transactions on Nuclear Science,* vol. 36, no. 6, pp. 2375-2382, 1989.

[32] S. Kuboyama *et al.*, "Single-event burnout of epitaxial bipolar transistors," *IEEE Transactions on Nuclear Science,* vol. 45, no. 6, pp. 2527-2533, 1998.

[33] A. Porzio *et al.*, "Experimental and 3D simulation study on the role of the parasitic BJT activation in SEB/SEGR of power MOSFET," in *Europian Radiations Effects on Components and Systems*, 2005.

[34] S. Liu *et al.*, "Single-event burnout and avalanche characteristics of power DMOSFETs," *IEEE Transactions on Nuclear Science,* vol. 53, no. 6, pp. 3379-3385, 2006.

[35] S. Liu *et al.*, "Effect of buffer layer on single-event burnout of power DMOSFETs," *IEEE Transactions on Nuclear Science,* vol. 54, no. 6, pp. 2554-2560, 2007.

[36] C. Dachs *et al.*, "Evidence of the ion's impact position effect on SEB in n-channel power MOSFETs," *IEEE Transactions on Nuclear Science,* vol. 41, no. 6, pp. 2167-2171, 1994.

[37] J. L. Titus *et al.*, "Experimental Studies of Single-Event Gate Rupture and Burnout in Vertical Power MOSFETs," *IEEE Transactions on Nuclear Science,* vol. 43, no. 2, pp. 533-545, 1996.

[38] G. H. Johnson *et al.*, "A Review of the Techniques Used for Modeling Single-Event Effects in Power MOSFETs," *IEEE Transactions on Nuclear Science,* vol. 43, no. 2, pp. 546-569, 1996.

[39] D. L. Oberg *et al.*, "First nondestructive measurements of power MOSFET single event burnout cross sections," *IEEE Transactions on Nuclear Science,* vol. 34, no. 6, pp. 1736-1741, 1987.

[40] T. A. Fischer *et al.*, "Heavy-ion-induced gate rupture in power MOSFETS," *IEEE Transactions on Nuclear Science,* vol. 34, no. 6, pp. 1786-1791, 1987.

[41] G. Soelkner *et al.*, "Charge carrier avalanche multiplication in high-voltage diodes triggered by ionizing radiation," *IEEE Transactions on Nuclear Science,* vol. 47, no. 6, pp. 2365-2372, 2000.

[42] C. Weiß *et al.*, "Numerical analysis of cosmic radiation-induced failures in power diodes," in *European Solid-state Devices and Circuits Conference,* 2011.

[43] M. V. O'Bryan *et al.*, "Compendium of single event effects for candidate spacecraft electronics for NASA," in *IEEE Radiation Effects Workshop,* 2012.

[44] J. L. Titus, "An updated perspective of single event gate rupture and single event burnout in power MOSFETs," *IEEE Transactions Nuclear Science,* vol. 60, no. 3, pp. 1912-1928, 2013.

[45] C. T. Kirk *et al.*, "A theory of transistor cut-off frequency (f ) falloff at high current densities," *IRE Transactions on Electron Devices,* vol. 9, no. 2, pp. 164-174, 1962.

[46] G. H. Johnson *et al.*, "Simulating single event burnout of N-channel power MOSFET's," *IEEE Transactions on Nuclear Science,* vol. 40, no. 5, pp. 1001-1008, 1993.

[47] A. Luu *et al.*, "Sensitive volume and triggering criteria of SEB in classic planar VDMOS," *IEEE Transactions on Nuclear Science,* vol. 57, no. 4, pp. 1900-1907, 2010.

[48] E. Mizuta *et al.*, "Investigation of Single-Event Damages on Silicon Carbide (SiC) Power MOSFETs," *IEEE Transactions on Nuclear Science,* vol. 61, no. 4, pp. 1924-1928, 2014.

[49] P. Li *et al.*, "Analysis of the influence of single event effects on the characteristics for SiC power MOSFETs," in *Prognostics System Health Management Conference,* 2017.

[50] J.-M. Lauenstein *et al.*, "Single-event effect testing of the CREE C4D40120D commercial 1200v silicon carbide Schottky diode," NASA Goddard SpaceFlight Center, 2014.

[51] J. A. McPherson *et al.*, "Simulation-based study of single-event burnout in 4H-SiC high-voltage vertical super junction DMOSFET: physical failure mechanism and robustness vs performance tradeoffs," *Applied Physics Letters,* vol. 120, no. 4, 2022.

[52] Y. Z. Cheng *et al.*, "4H-SiC UMOSFET With an Electric Field Modulation Region Below P-Body," *IEEE Transactions on Electron Devices,* vol. 67, no. 8, pp. 3298-3303, 2020.

[53] C. H. Yu *et al.*, "Simulation Study on Single-Event Burnout in Rated 1.2-kV 4H-SiC Super-Junction VDMOS," *IEEE Transactions on Electron Devices,* vol. 68, no. 10, pp. 5034-5040, 2021.

[54] D. A. Gajewski *et al.*, "SiC Power Device Reliability," in *International Integrated Reliability Workshop,* 2016.

[55] Q. Liu *et al.*, "Low on-resistance 1.2 kV 4H-SiC power MOSFET with Ron,sp of 3.4 m·cm2," *Journal of Semiconductors,* vol. 41, no. 6, p. 062801, 2020.

[56] D. R. Ball *et al.*, "Effects of Breakdown Voltage on Single-Event Burnout Tolerance of High-Voltage SiC Power MOSFETs," *IEEE Transactions on Nuclear Science,* vol. 68, no. 7, pp. 1430-1435, 2021.

[57] R. A. Johnson *et al.*, "Unifying Concepts for Ion-Induced Leakage Current Degradation in Silicon Carbide Schottky Power Diodes," *IEEE Transactions on Nuclear Science,* vol. 67, no. 1, pp. 135-139, 2020.

[58] J.-M. Lauenstein *et al.*, "Space radiation effects on SiC power device reliability," in *IEEE International Reliability Physics Symposium* , 2021.

[59] R. Baumann *et al.*, Radiation Handbook for Electronics, Texas Instruments, 2020.

[60] T. Tawara *et al.*, "Short minority carrier lifetimes in highly nitrogen doped 4H-SiC epilayers for suppression of the stacking fault formation in PiN diodes," *Journal of Applied Physics,* vol. 120, no. 11, 2016.

[61] T. Tawara *et al.*, "Suppression of the forward degradation in 4H-SiC PiN diodes by employing a recombination-enhanced buffer layer," *Material Science Forum,* vol. 897, pp. 419-422, 2017.

[62] J. Lu *et al.*, "Impact of Varied Buffer Layer Designs on Single-Event Response of 1.2-kV SiC Power MOSFETs," *IEEE Transactions on Electron Devices,* vol. 67, no. 9, pp. 3698-3704, 2020.

[63] F. Boige *et al.*, "Ensure an original and safe ‘fail-to-open’ mode in planar and trench power SiC MOSFET devices in extreme shortcircuit operation," *Microelectronics Reliability,* vol. 88, pp. 598-603, 2018.

[64] T. Nakamura *et al.*, "High performance SiC trench devices with ultra-low Ron," in *International Electron Devices Meeting*, 2011.

[65] X. Zhou *et al.*, "Single-Event Effects in SiC Double-Trench MOSFETs," *IEEE Transactions on Nuclear Science,* vol. 66, no. 11, pp. 2312-2318, 2019.

[66] C.-H. Yu *et al.*, "Research of Single-Event Burnout in Power Planar VDMOSFETs by Localized Carrier Lifetime Control," *IEEE Transactions on Nuclear Science,* vol. 62, no. 1, pp. 143-148, 2015.

[67] K. J. Bowman, "Advanced Sensors and Instrumentation Project Summaries," Idaho National Lab, 2021.

[68] N. Lophitis *et al.*, "TCAD Device Modelling and Simulation of Wide Bandgap Power Semiconductors," in *InTech*, 2018.

[69] D. Peters *et al.*, "1200V SiC Trench-MOSFET optimized for high reliability and high performance," in *European Conference on Silicon Carbide & Related Materials*, 2016.

[70] J. A. Felix *et al.*, "Enhanced Degradation in Power MOSFET Devices Due to Heavy Ion Irradiation," *IEEE Transactions on Nuclear Science,* vol. 54, no. 6, pp. 2181-2189, 2007.

[71] K. F. Galloway, "A Brief Review of Heavy-Ion Radiation Degradation and Failure of Silicon UMOS Power Transistors," *Electronics,* vol. 3, no. 4, pp. 582-593, 2014.

[72] S. Kuboyama *et al.*, "Characterization of Microdose Damage Caused by Single Heavy Ion Observed in Trench Type Power MOSFETs," *IEEE Transactions on Nuclear Science,* vol. 57, no. 6, pp. 3257-3261, 2010.

[73] S. Liu *et al.*, "Vulnerable Trench power MOSFETs under heavy ion irradiation," in *IEEE Nuclear and Space Radiation Effects Conference*, 2008.

[74] M. R. Shaneyfelt *et al.*, "Enhanced proton and neutron induced degradation and its impact on hardness assurance testing," *IEEE Transactions on Nuclear Science,* vol. 55, no. 6, pp. 3096-3105, 2008.

[75] G. I. Zebrev *et al.*, "Microdose induced drain leakage effects in power trench MOSFETs: Experiment and modeling," *IEEE Transactions on Nuclear Science,* vol. 61, no. 4, pp. 1531-1536, 2014.

[76] X. Wan *et al.*, "Charge deposition model for investigating SE-microdose effect in trench power MOSFETs," *Journal of Semiconductors,* vol. 36, no. 5, 2015.

[77] J.-M. Lauenstein *et al.*, "Recent Radiation Test Results for Trench Power MOSFETs," in *IEEE Radiation Effects Data Workshop*, 2017.

[78] Y. Wang *et al.*, "Simulation study of 4H-SiC UMOSFET structure with p+−polySi/SiC shielded region," *IEEE Transactions on Electron Devices,* vol. 64, no. 9, pp. 3719-3724, 2017.

[79] Y. Wang *et al.*, "Single-Event Burnout Hardness for the 4H-SiC Trench-Gate MOSFETs Based on the Multi-Island Buffer Layer," *IEEE Transactions on Electron Devices,* vol. 66, no. 10, pp. 4264-4272, 2019.

[80] Y. Wang *et al.*, "A Comparative Study of Single-Event-Burnout for 4H-SiC UMOSFET," *IEEE Transactions on Electron Devices,* vol. 10, pp. 373-378, 2022.

[81] J. Bi *et al.*, "Single-Event Burnout Hardening Method and Evaluation in SiC Power MOSFET devices," *IEEE Transactions on Electron Devices,* vol. 67, no. 10, pp. 4340-4345, 2020.

[82] J. Kim *et al.*, "Single-Event Burnout Hardening 4H-SiC UMOSFET Structure," *IEEE Transactions on Device and Materials Reliability,* vol. 22, no. 2, pp. 164-168, 2022.

[83] H. Tanaka *et al.*, "Ultra-low von and high voltage 4H-SiC heterojunction diode," in *IEEE 17th International Symposium of Power Semiconductor ICs,* 2005.

[84] O. Yasuhiko *et al.*, "Numerical analysis of specific on-resistance for trench gate superjunction MOSFETs," *Japanese Journal of Applied Physics,* vol. 54, no. 2, 2015.

[85] J. B. Baliga *et al.*, Fundamentals of power semiconductor devices, Springer Science & Business Media, 2010.

[86] S. Huang *et al.*, "Analysis of SEB and SEGR in super-junction MOSFETs," *IEEE Transactions on Nuclear Science,* vol. 47, no. 6, pp. 2640-2647, 2000.

[87] N. Ikeda *et al.*, "Single-event burnout of Super-junction power MOSFETs," *IEEE Transactions on Nuclear Science,* vol. 51, no. 6, pp. 3332-3335, 2004.

[88] L. Wang *et al.*, "A Novel Super-Junction Structure to Improve SEB Performance," in *14th IEEE International Conference on Solid-State and Integrated Circuit Technology*, 2018.

[89] K. Muthuseenu *et al.*, "Analysis of SEGR in Silicon Planar Gate Super-Junction Power MOSFETs," *IEEE Transactions on Nuclear Science,* vol. 68, no. 5, pp. 611-616, 2021.

[90] S. Katoh *et al.*, "Temperature dependence of single-event burnout for super junction MOSFET," in *International Symposium on Power Semiconductor Devices & IC's,* 2015.

[91] M. Zerarka *et al.*, "Analysis study of sensitive volume and triggering criteria of singleevent burnout in super-junction metal-oxide semiconductor field-effect transistors," *IET Circuits, Devices and Systems*, vol. 8, no. 3, pp. 197-204, 2014.

[92] K. Muthuseenu *et al.*, "Single-Event Gate Rupture Hardened Structure for High-Voltage Super-Junction Power MOSFETs," *IEEE Transactions on Electron Devices*, vol. 68, no. 8, pp. 4004-4009, 2021.

[93] S. Harada *et al.*, "First Demonstration of Dynamic Characteristics for SiC Superjunction MOSFET Realized using Multi-epitaxial Growth Method," in *IEEE International Electron Devices Meeting*, 2018.

[94] T. Masuda *et al.*, ".97mΩ-cm2/820V 4H-SiC super junction V-groove trench MOSFET," *Material Science Forum*, vol. 897, pp. 483-488, 2017.

[95] X. Zhou *et al.*, "Performance limits of vertical 4H-SiC and 2H-GaN superjunction devices," *Material Science Forum*, vol. 963, pp. 693-696, 2019.

[96] J. A. McPherson *et al.*, "Robustness of Semi-Superjunction 4H-SiC Power DMOSFETs to Single-Event Burnout from Heavy Ion Bombardment," *Materials Science Forum*, vol. 1062, pp. 683-687, 2022.

[97] J. A. McPherson *et al.*, "Heavy Ion Transport Modeling for Single-Event Burnout in SiC-Based Power Devices," *IEEE Transactions on Nuclear Science*, vol. 66, no. 1, pp. 474-481, 2019.